\documentclass[aps,prd,nopacs,floatfix,notitlepage,nofootinbib,twocolumn,a4paper,longbibliography]{revtex4-1}

\usepackage{amsfonts,amsmath,units,wasysym,epsfig,graphicx,verbatim,color,subfigure,graphicx,bm,mathrsfs,lipsum,hyperref,cleveref}
\usepackage{booktabs}
\usepackage[normalem]{ulem}  % \sout{old text} for strikeout
\usepackage{multirow}
\usepackage{xcolor}

\DeclareUnicodeCharacter{2032}{'}
\DeclareUnicodeCharacter{2033}{''}

\begin{document}

\newcommand{\bhaskar}[1]{\textcolor{blue}{ \bf BB: #1}}
\newcommand{\OKC}{The Oskar Klein Centre, Department of Astronomy, Stockholm University, AlbaNova, SE-10691 Stockholm, Sweden}
\newcommand{\HU}{{Hamburger Sternwarte, Gojenbergsweg 112, D-21029 Hamburg, Germany}}
\newcommand{\LSUC}{{Center for Computation \& Technology, Louisiana State University, 70803, Baton Rouge, LA, USA}}
\newcommand{\LSUD}{{Department of Physics \& Astronomy, Louisiana State University, 70803, Baton Rouge, LA, USA}}
\newcommand{\LS}{\textbf{\textcolor{orange}{LS:}}}
\newcommand{\lscomment}[1]{\textcolor{orange!90!white}{#1}}

\title{Binary neutron star mergers with SPHINCS\_BSSN: temperature-dependent equations of state and damping of
constraint violations}

\author{Bhaskar Biswas$^{\rm 1, \rm 2}$, Stephan Rosswog$^{\rm 1, \rm 2}$, Peter Diener$^{\rm 3, \rm 4}$, Lukas Schnabel$^{\rm 1}$}
\affiliation{$^{\rm 1}$\HU, $^{\rm 2}$\OKC,$^{\rm 3}$\LSUC, $^{\rm 4}$\LSUD}

% own commands
\newcommand{\nifs}{\ensuremath{^{56}\mathrm{Ni}}}
\newcommand{\Nifs}{\ensuremath{^{56}\mathrm{Ni}} $\;$}
\def\paren#1{\left( #1 \right)}
\def\Mesz{M\'esz\'aros~}
\def\Pacz{Paczy\'nski~}
\def\Kluz{Klu\'zniak~}
\def\p{\partial}
\def\msun{M$_{\odot}$}
\def\Msun{M$_{\odot}$ }
\def\be{\begin{equation}}
\def\ee{\end{equation}}
\def\bi{\begin{itemize}}
\def\i{\item}
\def\ei{\end{itemize}}
\def\ben{\begin{enumerate}}
\def\een{\end{enumerate}}
\def\bea{\begin{eqnarray}}
\def\eea{\end{eqnarray}}
\def\bt{\begin{tabbing}}
\def\et{\end{tabbing}}
\def\gcc{gcm$^{-3}$}
\def\Gcc{gcm$^{-3} \;$}
\def\ccm{cm$^3$}
\def\edo{\end{document}}
\def\red{\textcolor{red}}
\def\blue{\textcolor{blue}}
\def\green{\textcolor{green}}
\def\ye{$Y_e$}
\def\Ye{$Y_e$ }
\def\rs{$R_{\rm S}$}
\def\Rs{$R_{\rm S}$ }
\def\noin{\noindent}
\def\QIone{\mathcal{Q}_{\; \; 1}}
\def\QItwo{\mathcal{Q}_{\; \; 2}}
\def\QIthree{\mathcal{Q}_{\; \; 3}}
\def\QIfour{\mathcal{Q}_{\; \; 4}}
\newcommand{\ud}{\mathrm{d}}
\newcommand{\sn}{\ensuremath{\mathrm{sn}}}
\newcommand{\cn}{\ensuremath{\mathrm{cn}}}
\newcommand{\dn}{\ensuremath{\mathrm{dn}}}
\newcommand{\sech}{\ensuremath{\mathrm{sech}}}
\newcommand{\nd}{\noindent}
\newcommand{\enth}{\mathcal{E}}
\newcommand{\carb}{$^{12}$C}
\newcommand{\Carb}{$^{12}$C $\;$}
\newcommand{\ox}{$^{16}$O}
\newcommand{\Ox}{$^{16}$O $\;$}
\newcommand{\tlg}{\tilde{\gamma}}
\newcommand{\tlG}{\tilde{\Gamma}}
\newcommand{\emfp}{e^{-4\phi}}
\newcommand{\tlA}{\tilde{A}}
\newcommand{\tlR}{\tilde{R}}
\newcommand{\xt}{\tilde{x}}
\newcommand{\dt}[1]{\partial_t #1}
\newcommand{\pdu}[3]{\partial_{#3} #1^{#2}}
\newcommand{\pdl}[3]{\partial_{#3} #1_{#2}}
\newcommand{\pdpdu}[4]{\partial_{#3}\partial_{#4} #1^{#2}}
\newcommand{\pdpdl}[4]{\partial_{#3}\partial_{#4} #1_{#2}}
\newcommand{\upwindu}[3]{\beta^{#3}\bar{\partial}_{#3} #1^{#2}}
\newcommand{\upwindl}[3]{\beta^{#3}\bar{\partial}_{#3} #1_{#2}}
\newcommand{\tlGn}{\tilde{\Gamma}_{\mathrm{(n)}}}
\newcommand{\tlGmixed}[2]{\tlG_{#1}^{\;\;\;#2}}

\def\rg{$\vec{r}_{\rm G}$}
\def\ra{$\vec{r}_a$}
\def\rb{$\vec{r}_b$}
\def\peak{\rm peak}
\def\ji{\varphi}
\def\rin{\mathcal{I}}
\def\x{{\bf x}}
\def\s{{\bf s}}
\def\k{{\bf k}}
\def\n{{\bf n}}
\def\f{{\bf f}}
\def\pd{\partial}
\def\fa{f_{\alpha}(\x,\p,t)}
\def\kabs{\kappa^{\rm abs}_\nu}
\def\ksca{\kappa^{\rm sca}_\nu}
\def\Flux{{\bf F}_\nu}
\def\Frad{{\bf F}_{\rm rad}}
\def\Pressure{{\mathsf P}_\nu}
\def\kB{k_B}
\def\tauph{\tau_{\it ph}}
\def\taumu{\tau_\mu}
\def\MSun{M_{\astrosun}}

% Journals
%\newcommand{\apj}{Ap. J., }
\newcommand{\apjl}{Ap. J. Lett., }
\newcommand{\apjs}{Ap. J. Supp., }
\newcommand{\physrep}{Phys. Rep., }
\newcommand{\mnras}{Mon. not. RAS., }
\newcommand{\apss}{Astroph. \& Space Sci., }
\newcommand{\aap}{Astron. \& Astrophys., }
\newcommand{\pasp}{Publications of the Astronomical Society of the Pacific,}

% SPH kernels
\def\M4{M$_4$}
\def\M6{M$_6$}
\def\lcm6{LCM$_6$}
\def\qcm6{QCM$_6$}
\def\whn{W$_{\rm H,n}$}
\def\wh3{W$_{\rm H,3}$}
\def\wh4{W$_{\rm H,4}$}
\def\wh5{W$_{\rm H,5}$}
\def\wh6{W$_{\rm H,6}$}
\def\wh7{W$_{\rm H,7}$}

% vector algebra
\def\divv{\nabla \cdot \vec{v}}
\def\vv{\vec{v}}
\def\dx{\Delta x}

% alternative vector symbole
\newcommand{\ve}[1]{\boldsymbol{#1}}

% turbulence
\def\ed{\dot{\epsilon}}
\def\vl{v_\lambda}
\def\vort{\vec{\mathcal{W}}}

% reduced quadrupole moment
\newcommand{\rqm}{{%
      \declareslashed{}{\text{-}}{0.04}{0}{I}\slashed{I}}}

% code names
\newcommand{\SpB}{\texttt{SPHINCS\_BSSN}\; }
\newcommand{\spB}{\texttt{SPHINCS\_BSSN}}
\newcommand{\spid}{\texttt{SPHINCS\_ID}}
\newcommand{\spiD}{\texttt{SPHINCS\_ID}\; }
\newcommand{\Ma}{\texttt{MAGMA2}\;}
\newcommand{\ma}{\texttt{MAGMA2}}
\newcommand{\Lo}{\texttt{LORENE}\;}
\newcommand{\lo}{\texttt{LORENE}}
\newcommand{\Fu}{\texttt{FUKA}\;}
\newcommand{\fu}{\texttt{FUKA}}
\newcommand{\comp}{\texttt{CompOSE}\;}
\newcommand{\binns}{\texttt{BinNS}\;}
\newcommand{\McL}{\texttt{McLachlan}\;}
\newcommand{\Mhdu}{\texttt{MHDUET}\;}
\newcommand{\mhdu}{\texttt{MHDUET}}
\newcommand{\Sam}{\texttt{SAMRAI}\;}
\newcommand{\sam}{\texttt{SAMRAI}}
\newcommand{\Simf}{\texttt{Simflowny}\;}
\newcommand{\simf}{\texttt{Simflowny}}

\begin{abstract}
Neutron star mergers hold the key to several grand challenges of contemporary
(astro-)physics. In view of the upcoming next generation of ground-based detectors,
it is crucial to keep improving theoretical predictions to harvest the full
scientific returns from these investments. We introduce here a substantial update of our
Lagrangian numerical relativity code \spB. Apart from changing our unit
system, we add constraint damping terms to the BSSN spacetime evolution equations.
We demonstrate that this measure reduces, without 
noteworthy computational cost, the Hamiltonian constraint violations by more 
than an order of magnitude. We further implement contributions to 
thermal energy and pressure that are based on Fermi liquid theory and contain a parametrization of the Dirac effective mass. These 
terms can be combined with any cold equation of state, and they enhance the 
physical realism of our simulations and  introduce a physics-based concept of a
temperature. In a set of merger simulations, we demonstrate good agreement 
with other temperature-dependent numerical relativity simulations. We find that
different parametrizations of the Dirac effective mass can translate into 
shifts of $\sim 150$ Hz in the dominant post-merger gravitational wave peak frequency.

\end{abstract}

\maketitle

\section{Introduction}
The merger process of two neutron stars is shaped by physics under the most extreme conditions, involving strongly
curved and dynamically changing spacetimes \cite{alcubierre08,baumgarte10,bona11,rezzolla13a,shibata16} with potential black hole (BH) formation, matter densities exceeding
several times nuclear matter density \cite{lattimer12a,baym18}, magnetic fields that are locally amplified to mind-boggling field strengths \cite{price06,anderson08b,rezzolla11,zrake13,kiuchi15,palenzuela22,aguilera22,aguilera24,kiuchi24,aguilera25}, and neutrino emission at luminosities that, for short times, exceed the solar photon luminosity by  20 orders of magnitude \cite{rosswog03a,dessart09,sekiguchi16a,radice18a,foucart23}.

Although the extreme-physics environment is interesting in its own right, neutron star mergers
are also closely related to many other grand challenges of contemporary physics and astrophysics,
including dark matter questions \cite{das21,fortin21,emma22,baryakhtar22}, deviations from GR \cite{barack19,Biswas:2023ceq}, or the properties of hot and dense nuclear matter \cite{lattimer16,abbott18b,Biswas:2024hja} that may leave imprints on the post-merger 
GW signal \cite{baiotti19}. Neutron star mergers have been suggested as sources of heavy elements for a long time
\cite{symbalisty82,eichler89,rosswog99,freiburghaus99b} and, as of today, they are the only confirmed cosmic source of r-process elements \cite{tanvir17,kasen17,arcavi17,rosswog18a,metzger19a}. Moreover, neutron star mergers can power short and possibly
even long gamma ray bursts \cite{gehrels06,zhang07,troja22,rastinejad22,yang24,levan24} and they  are ``standard sirens" that offer an alternative Hubble
parameter estimation \cite{schutz86,holz05,nissanke10}.

To make significant progress on these challenging questions requires the comparison
of signatures of several messengers, ideally gravitational and electromagnetic waves and neutrinos,
with theoretical models. Thus, there is a strong impetus to continuously improve on the physics and
numerics of neutron star merger models.

While numerical relativity is traditionally performed with Eulerian methods \cite{alcubierre08,baumgarte10,bona11,rezzolla13a,shibata16}, we  have recently begun 
to develop an alternative methodology where we use well-established mesh-based methods for the evolution of the  spacetime, but we use  Lagrangian particles to evolve the fluid \cite{monaghan05,rosswog09b,rosswog15c}. 
This methodology is implemented in the \SpB code \cite{rosswog21a}. The ejecta of neutron star mergers are
only about $\sim$ 1\% of the binary mass, but their exact properties are 
crucial for multi-messenger astrophysics, since they shape the whole electromagnetic emission.
Lagrangian methods have major advantages when it comes to following merger ejecta: advection 
is exact (i.e. a particle carries matter properties such as the 
electron fraction without any loss of information 
while in a Eulerian method this is subject to resolution), and the particle nature provides
naturally ``fluid tracers" without any need for additional computational infrastructure. 
Moreover, the particle motion is not restricted by grid boundaries and can, in principle, be 
followed out to infinity. As an example, in the (essentially Newtonian) simulations of 
\cite{rosswog14a} the ejecta were followed up to 100 years post-merger.

The \SpB code \cite{rosswog21a} has been written from scratch and in its original version solved the BSSN equations \cite{alcubierre08,baumgarte10,bona11,rezzolla13a,shibata16} on a uniform mesh and used a polytropic equation
of state. These ingredients allowed for a comparison with the tests that were performed 
in the pioneering Eulerian
numerical relativity simulations. We found generally excellent agreement with results from
the literature, for example, for the challenging ``migration test" where an unstable neutron star, depending on the initial perturbation, transitions \cite{font02,cordero09,bernuzzi10},
 either 1) to a wildly oscillating neutron star or 2) it collapses into a BH. We also found excellent agreement with the literature in terms of the frequencies of oscillating neutron stars \cite{rosswog21a,rosswog25c}. In the latter publication, we found that the first five harmonics of an oscillating neutron star agreed to sub-percent precision with the results from perturbation codes.

In the first \SpB implementation, the energy momentum tensor 
at the particle positions was mapped to the spacetime mesh via high-accuracy  kernels borrowed from vortex methods \cite{cottet00}.
In the first paper discussing neutron star merger simulations with \SpB \cite{diener22a} we used a (fixed in time) mesh-refinement. In this paper we further improved the accuracy of the mapping between the particles and the mesh, and we translated the ``artificial
pressure method" \cite{rosswog20a}, originally designed to create high-accuracy Newtonian  SPH initial conditions, to general relativity. 
In a companion paper \cite{rosswog22b} we introduced piecewise polytropic approximations to nuclear
physics-based equations of state in \SpB and we approximated thermal effects as an ideal-gas law
with a thermal exponent $\Gamma_{\rm th}$. We applied this new methodology to simulate binary neutron star
mergers with different equations of state and we explored their gravitational wave emission and  the impact of the thermal exponent
$\Gamma_{\rm th}$, e.g.\ on the dynamically ejected matter. 

The next major code update, \texttt{SPHINCS\_BSSN\_v1.0},
is documented in \cite{rosswog23a}. It contains a further improvement
of the particle-to-mesh mapping via a local
regression estimate (LRE) coupled with a ``multi-dimensional optimal order detection" (MOOD) algorithm and
it allows dynamical refinement of the spacetime mesh
 to robustly follow neutron stars that collapse into
BHs. The astrophysical applications of \texttt{SPHINCS\_BSSN\_v1.0} so far contain the study of the
multi-messenger signatures of neutron star mergers where only one of the two stars is rapidly spinning
\cite{rosswog24b}. This scenario should be realized in $\sim$ 5\% of the merging binaries. Another application was the
exploration of fast ejecta in a neutron star merger \cite{rosswog25b}, where, enabled by the Lagrangian nature
of \spB, we identified two different mechanisms to eject semi-relativistic matter ($v > 0.5c$).

In the present paper, we describe three new developments in \spB. First, we give up on our original convention to measure
all energies in terms of the nucleon rest mass energy $m_0 c^2$. While this convention makes the basic hydrodynamic
equations look very clean and in particular makes the baryon number conservation very transparent, it makes it difficult to add new physics because of 
the somewhat involved units. Second, we improve our spacetime evolution equations.
So far we have followed the standard BSSN approach, where one starts from
constraint-satisfying initial conditions and then evolves the BSSN variables forward in time and monitors
possible constraint violations, but without trying to suppress such violations. Now, we add additional constraint-damping
terms to our BSSN formulation where we closely follow the suggestions that Etienne \cite{etienne24} made in the Eulerian
context. Last but not least, we improve on our treatment of thermal effects in the equation of state.
Instead of using a fixed thermal $\Gamma_{\rm th}$, we use contributions to pressure and thermal energy 
that are based on Fermi liquid theory, e.g. \cite{constantinou14,constantinou15,lattimer16}. Here we follow closely the work of Raithel et al. 
\cite{Raithel:2019gws,raithel21a} that have followed such an approach in the Eulerian
context. Apart from enhancing the physical realism
of the equation of state, this approach also 
provides a physics-based temperature that is needed, for example, for neutrino reactions.

Our paper is structured as follows. In the first part of Sec.~\ref{sec:methodology} we concisely summarize
the hydrodynamics equations, now {\em without} the convention of measuring all energies in units of
$m_0 c^2$. In the second part, we summarize our constraint damping procedure and in the third part the
thermal equation of state (EOS), that can be combined with any cold matter EOS, is explained in detail. In
Sec.\ref{sec:results}, we  demonstrate the effects of constraint damping and of the thermal EOS
before we conclude in Sec.~\ref{sec:summary}.

\section{Improved Methodology}
\label{sec:methodology}
This paper introduces three new elements in the \SpB simulation methodology:  
new conventions in formulating the equations, additional constraint damping 
terms for the spacetime evolution and physics-based thermal contributions 
in the equation of state that are based on Fermi liquid theory \cite{baym91}.

\subsection{General relativistic Hydrodynamics with SPHINCS\_BSSN}
\label{sec:hydro}
In our previous work related to \SpB we have always followed the convention to 
measure all forms of energy in terms of the nucleon rest mass energy $m_0 c^2$, 
since this makes the equations very elegant and the conservation of nucleon 
number particularly transparent. The resulting units, however, are somewhat 
cumbersome. Therefore, we give up on this convention here, so that our new units 
are exactly the same as the standard units used in numerical relativity codes with $G=c=1$. The 
derivation of the SPH  equations can be found in explicit detail in 
\cite{rosswog09b,rosswog10a} and, in more concise form in 
\cite{monaghan01,rosswog15c,rosswog25c}.

Explicitly, the changes are the following: what we so far called ``baryon 
number density measured in the computing frame", $N$, now becomes mass density 
in the computing frame, $\rho^\ast= m_0 N$, where $m_0$ is the baryon 
mass\footnote{In principle, the average baryonic mass depends on the exact 
composition. In practice, however, deviations from the atomic mass unit are 
even in the worst case much smaller than 1\%, see Sec. 2.1.1 in \cite{diener22a}.
Therefore, we use the numerical value of the atomic mass unit for $m_0$.} and what 
we, so far, used as ``pressure" was actually pressure in units of $m_0 c^2$ (since 
it has dimensions of an energy density). Giving up our previous $m_0 c^2$ 
convention and ignoring the grad-h terms (derived for the general relativistic 
case in \cite{rosswog10a}),  our GR SPH equations in a fixed computing frame 
read 
\bea
\rho^\ast_a &=& \hspace*{0.3cm}\sum_b m_b \; W_{ab}(h_a) \\
\left(\frac{dS_i}{dt} \right)_a &=& \left(\frac{d(S_i)_a}{dt} \right)_{\rm hyd} +\left(\frac{d(S_i)_a}{dt} \right)_{\rm met} \\
\frac{de_a}{dt}  &=& \left(\frac{de_a}{dt} \right)_{\rm hyd} +\left(\frac{de_a}{dt} \right)_{\rm met},
\eea
where the hydrodynamic contributions read
\bea
\left(\frac{d(S_i)_a}{dt} \right)_{\rm hyd} &=& -  \sum_b m_b \left[\frac{P_a}{\rho_a^{\ast 2}}  D^a_i  +  \frac{P_b}{\rho_b^{\ast 2}} D^b_i \right]\\
\left(\frac{de_a}{dt} \right)_{\rm hyd} &=& -  \sum_b m_b \left[\frac{P_a v_b^i}{\rho_a^{\ast 2}}  D^a_i  +  \frac{P_b v_a^i}{\rho_b^{\ast 2}} D^b_i \right],
\eea
with
\be
D^a_i \equiv \frac{\sqrt{-g}_a \p W_{ab}(h_a)}{\p x_a^i} \; {\rm and} \; D^b_i \equiv \frac{\sqrt{-g}_b \p W_{ab}(h_b)}{\p x_a^i}.
\ee
The metric terms read
\bea
\left(\frac{d(S_i)_a}{dt} \right)_{\rm met} = \left( \frac{\sqrt{-g}}{2 \rho^\ast} T^{\mu \nu} \frac{\p g_{\mu \nu}}{\p x^i}\right)_a
\eea
and 
\bea
\left(\frac{de_a}{dt} \right)_{\rm met} = -\left( \frac{\sqrt{-g}}{2 \rho^\ast} T^{\mu \nu} \frac{\p g_{\mu \nu}}{\p t}\right)_a.
\eea
Note that we have made the {\em choice} to calculate $\rho^\ast$ via a summation approach, 
because the combination of positive masses and positive definite SPH kernels $W$
is guaranteed to deliver a positive density estimate. One could, however, also integrate 
$\rho^\ast$ forward in time as an alternative. Our computing frame density $\rho^\ast$ is 
related to the local rest frame density $\rho$ via
\be
\rho^\ast= \sqrt{-g} \Theta \rho,
\ee
where $\Theta$ is the generalized Lorentz factor
\be
\Theta= \frac{1}{\sqrt{-g_{\mu\nu} v^\mu v^\nu}}
\ee
expressed in terms of the coordinate velocities $v^\mu= dx^\mu/dt$.
Note also that, for simplicity, we here use the same symbols $S$ and $e$ as in previous work, although now they
have different units. To be able to robustly
treat shocks, one has to add dissipative
terms. This is done as described in Sec. 2.1.1 of \cite{rosswog22b}.

\subsection{Constraint damping}
\label{sec:constraint_damping}
To harvest the full scientific potential of the next generation of gravitational wave detectors, numerical relativity simulations need to become more accurate \cite{puerrer20,ferguson21}. 
One obvious way forward is increased numerical resolution, the other is to improve the
methods so that at the same resolution higher accuracy can be achieved. In the spirit of the second approach, we improve our spacetime evolution in \spB, by closely following the suggestions made by Etienne~\cite{etienne24} in 
the Eulerian context.\\
The first change is a modification to the strength of the Kreiss-Oliger (KO)~\cite{Kreiss1973MethodsFT} dissipation for various variables in
different regions. The idea is that dissipation should be decreased where the curvature is large, since we 
do not want to flatten out large gradients in the fields and dissipation 
should be increased for the gauge variables, since we want to smooth out sharp features as much as possible. It is convenient to use
the conformal factor $\phi$ as a measure of the curvature and the strength of the dissipation then becomes dependent on $\phi$ as
\begin{equation}
    \epsilon_{\mathrm{KO}}(\phi) = \epsilon_{\mathrm{KO,CA}} e^{-2\phi},
\end{equation}
where $\epsilon_{\mathrm{KO,CA}}$ is 0.99 for gauge variables and 0.2 for all other BSSN variables. This modification is expected to be less important for
binary neutron stars compared to binary black hole systems as the curvature is smaller, and features
in the gauges are expected to be less sharp.

The second improvement is the addition of constraint damping, and therefore involves the Hamiltonian constraint, which in terms of BSSN variables can be written as
\begin{equation}
{\cal H}=\frac{2}{3}K^2-\tilde{A}_{ij}\tilde{A}^{ij}+e^{-4\phi}(\tilde{R}-8\tilde{D}^i\phi\tilde{D}_i\phi-8\tilde{D}^2\phi)=0.
\end{equation}
The needed change is an additional term to the RHS equation for the conformal factor $\phi$, i.e.\
\begin{equation}
    \partial_t\phi =[\partial_t\phi]_{\mathrm{STD}}-\kappa_{\phi{\cal H}} {\cal H},
\end{equation}
where $\kappa_{\phi\cal H}$ has to be chosen small enough to maintain
stability. As the constraints contain second derivatives of the field variables, this essentially 
adds a parabolic term to the evolution equation for $\phi$. This introduces a Courant stability requirement, see e.g.~\cite{press92},
 such that the timestep $\Delta t$ scales as $\Delta s_n^2$, where $\Delta s_n$ is the spatial resolution on refinement level $n$. 
That means that $\kappa_{\phi{\cal H}}$ has to be small where the resolution is high, but can be increased significantly where the resolution is low. Since we are taking the same timestep on all refinement levels, we choose to scale $\kappa_{\phi{\cal H}}$ as
\begin{equation}
    \kappa_{\phi{\cal H}}= C\frac{\Delta s_{n}^2}{\Delta t}.
    \label{eq:cdamp}
\end{equation}
In~\cite{etienne24} it was reported, based on many experiments in a Eulerian context, that the 
value $C=0.15$ leads to stable evolutions. We have found problems in some simulations when 
using this value of $C$, so we decrease this value slightly to $C=0.14$, which has worked well in all cases so far.

\subsection{Thermal equation of state}
\label{sec:th_EOS}
Much of the merger dynamics and final outcome (e.g.\ whether/when a BH forms) --
as well as the observable signatures (GWs, neutrinos and ejecta/electromagnetic emission) -- crucially depend on the (not so well known) nuclear matter equation of
state (EOS). The simplest approximation is a polytropic EOS with a 
single polytropic exponent $\Gamma$. A significantly better approach is to 
use piecewise polytropic EOSs whose polytropic pieces have been fitted to cold 
nuclear matter EOSs, e.g. \cite{read09}. Such approaches are often enhanced by 
an ideal gas law type thermal EOSs that approximate thermal 
effects with a thermal exponent $\Gamma_{\rm th}$.\\
Clearly better, however, are EOSs that are based on the state-of-the-art 
microscopic calculations. Such calculations are way to expensive to be 
performed ``on the fly" during a dynamical merger simulation. Therefore, 
the most common approach is to interpolate between values  in pre-computed tables, provided 
for example via the COMPOSE database \cite{compose}.
The use of tabulated EOS requires non-trivial algorithms to recover 
the physical (``primitive") variables (e.g. $v^i, P, u$) from the numerical 
(``conservative") ones ($\rho^\ast, S_i, e$, see Sec.~\ref{sec:hydro})
and we are also following such an approach within \SpB \cite{shankar26}.
Once the ``con-to-prim" algorithm is working robustly, it is straight forward 
to explore new EOS-physics: one only needs to replace one table by another. \\
But there are also shortcomings related to tabulated equations of state. First, the
``con-to-prim" recovery usually requires many table interpolation calls that come
at a high computational burden. Moreover, such tables can be non-smooth which
may require many iterations to convergence, and, in the worst case, one may 
even fail to find the correct physical solution. In such cases one may still try to find a solution by means of a more robust, but usually slower
fallback-algorithm. Second, the tables are usually restricted 
to temperatures above $\sim$ 0.1 MeV and to densities above $\sim 10^3$ \gcc.
If the focus of a study are the ejecta of a merger, these limits may become
detrimental, since the ejecta very quickly run outside of the table boundaries and
one has to resort to fallback strategies such as calculating pressures etc. from the
(too high) table boundaries, blending the EOS with another, lower-density EOS or
even extrapolating, none of which is an ideal option.\\
We adopt here the finite-temperature extension introduced by
Raithel et al.~\cite{Raithel:2019gws}, which provides a physically
motivated and analytic prescription for augmenting cold equations of
state with thermal effects. In the following, we summarize the main
elements of this framework and describe our implementation in \spB.

\subsubsection{Finite-temperature EOS framework}
Following Ref.~\cite{Raithel:2019gws}, the energy per particle of nuclear
matter is expanded as a Taylor series in the neutron excess
parameter $(1-2Y_p)$ up to second order:
\begin{equation}
    E_{\text{nuc}} (n,Y_p,T) = E_{\text{nuc}} (n,Y_p=1/2,T) + E_{\text{sym}} (n,T)(1-2Y_p)^2.
    \label{eq:Enuc}
\end{equation}
Here, $E_{\text{nuc}}(n, Y_p=1/2, T )$ represents the energy of symmetric nuclear matter. The symmetry energy is defined as
\begin{equation}
    E_{\text{sym}}(n,T) = \frac{1}{2} \frac{\partial^2 E_{\text{nuc}} (n,Y_p,T)}{\partial (1-2Y_p)^2} \Big|_{Y_p =1/2}.
\end{equation}
To satisfy charge neutrality, we set the proton and electron densities equal, leading to
\begin{equation}
    n_e = Y_p n.
\end{equation}
Expanding Eq.~(\ref{eq:Enuc}), the energy of symmetric nuclear matter can be expressed as the sum of the cold symmetric matter energy and the thermal contribution:
\begin{align}
    E_{\text{nuc}} (n,Y_p,T) &= E_{\text{nuc}} (n,Y_p=1/2,T=0)  \notag 
    \\
    &\quad + E_{\text{nuc,th}} (n,Y_p=1/2,T) \notag \\
    &\quad + E_{\text{sym}} (n,T)(1-2Y_p)^2.
\end{align}
Since the model is based on a cold EOS in $\beta$-equilibrium, this should be reflected in the energy per particle. To achieve this, we eliminate the cold symmetric term as,
\begin{align}
    E_{\text{nuc}} (n,Y_p,T) &= E_{\text{nuc}} (n,Y_{p,\beta},T=0) + E_{\text{nuc,th}} (n,1/2,T) \notag \\
    &\quad + E_{\text{sym}} (n,T)(1-2Y_p)^2 \notag \\
    &\quad - E_{\text{sym}} (n,T=0)(1-2Y_{p,\beta})^2.
\end{align}
Here, $Y_{p,\beta}$ denotes the proton fraction at zero temperature in $\beta$-equilibrium. \\
The final step in completing this generalized model is to incorporate the contributions of leptons and photons into the system. These contributions can also be written as the sum of a cold  and a thermal part. In the zero-temperature limit, the energy contribution from relativistic degenerate electrons is given by  
\begin{equation}
E_{\text{lepton}} (n,Y_p,T=0) = 3K Y_p (Y_p n)^{1/3},
\end{equation}
where
\begin{equation}
K \equiv \frac{(3\pi^2)^{1/3}}{4} \; \hbar c.
\end{equation}
The total energy is given by:
\begin{align}
    E(n,Y_p,T) &= E(n,Y_p,T=0) + E_{\text{th}}(n,Y_p,T), \notag \\
    E(n,Y_p,T=0) &= E(n,Y_{p,\beta},T=0) + E_{\text{sym}}(n,T=0) \notag \\
    &\quad \times [(1-2Y_p)^2 - (1-2Y_{p,\beta})^2] \notag \\
    &\quad + 3K \left(Y_p^{4/3} - Y_{p,\beta}^{4/3} \right) n^{1/3}, \notag \\
    E_{\rm th}(n,Y_p,T) &= E_{\text{nuc,th}} (n,1/2,T) \notag \\
    &\quad + E_{\text{lepton, th}} (n,Y_p,T) \notag \\
    &\quad + E_{\text{sym}} (n,T)(1-2Y_p)^2
\end{align}

\subsubsection{Symmetry energy at zero temperature} The symmetry-energy parameterization adopted here follows exactly the
prescription of Li et al.~\cite{Li:2014vua} as implemented in
Ref.~\cite{Raithel:2019gws},
\begin{equation}
\label{eq:Esym}
E_{\rm sym}(n, T=0)= \eta E_{\rm sym}^{\rm kin}(n)  
	+ \left[ S_0 - \eta E_{\rm sym}^{\rm kin}(n_{\rm sat}) \right] \left(\frac{n}{n_{\rm sat}}\right)^{\gamma}.
\end{equation}
Here, \( E_{\text{kin}}(n) \) represents the kinetic part of the symmetry energy, while the second term corresponds to the potential part. Since the expression for the potential symmetry energy is not well-constrained by experiments, it is written as the difference between the symmetry energy at saturation density, \( S_0 \equiv E_{\text{sym}}(n_{\text{sat}}) \), and the kinetic part. A density dependence is introduced through the parameter \( \gamma \), which controls the rate of change of the symmetry energy due to short-range correlations. Finally, the parameter \( \eta \) is introduced to account for short-range interactions between proton-neutron pairs.\\
The kinetic symmetry energy can be expressed in terms of the  momentum distribution. The kinetic energy per particle, \( \varepsilon_{k,q} \), is given by
\[
\varepsilon_{k,q} = 3E_f(n_q) n^{5/3},
\]
where \(q\) represents the particle (either a neutron or proton) and \( E_f(n) \) is the Fermi energy
\[
E_f(n_q) = \frac{\hbar^2}{2m} \left( 3 \pi^2 n_q \right)^{2/3},
\]
with \( m \) representing the mass of the particle. The small difference between the proton and neutron masses is neglected, so we approximate \( m \approx m_n \), where \( m_n \) is the neutron mass.\\
Since the kinetic symmetry energy arises due to the change in the Fermi energy of a gas as the proton fraction varies between symmetric matter and neutron matter, it can be expressed as a function of the total baryon density \( n \) as
\begin{align}
\begin{split}
\label{eq:Ekinsym}
E_{\rm sym}^{\rm kin}(n)  &= \frac{3}{5} \left[ 2E_f\left(n_p = n_n =\frac{1}{2} n\right) - E_f(n_n = n)  \right]  \\
 &= \frac{3}{5} \left(2^{1/3} -1 \right)  E_f(n),
\end{split}
\end{align}
where \( E_f(n) \) represents the Fermi energy as a function of total density.\\
An expression for the symmetry energy at zero temperature can be derived, see e.g. \cite{blaschke16,raithel20}, which depends on three parameters: \( S_0 \), \( L \), and \( \gamma \), all of which are experimentally constrained. The remaining task is to establish a relationship between the symmetry energy at zero temperature and the proton fraction for matter in beta-equilibrium, \( Y_{p,\beta} \), which is the primary result extracted from our data. This relationship is given by
\begin{equation}
\label{eq:YpBInv}
Y_{p,\beta} = \frac{1}{2} + \frac{(2 \pi^2)^{1/3}}{32} \frac{n}{\xi} \left\{ (2\pi^2)^{1/3} - \frac{\xi^2}{n} \left[\frac{\hbar c}{E_{\rm sym}(n,T=0)}\right]^3 \right\},
\end{equation}
where, for convenience, one introduces the auxiliary quantity \( \xi \), defined as
\begin{multline}
\label{eq:xi}
\xi \equiv \left[ \frac{E_{\rm sym}(n,T=0)}{\hbar c} \right]^2 \times \\
\left\{ 24 n \left[ 1+ \sqrt{ 1 +  \frac{\pi^2 n}{288}\left(\frac{\hbar c}{E_{\rm sym}(n,T=0)} \right)^3}\right]  \right\}^{1/3}.
\end{multline}
At densities below \( 0.5 \, n_s \), nuclei begin to form, and the expansion formalism for the nuclear symmetry energy breaks down. To handle this transition, we require a reasonable extrapolation of the symmetry energy model to integrate with a known low-density EoS table. Following Ref.~\cite{Most:2021ktk}, we empirically choose a power-law decay model for the symmetry energy extrapolation to ensure that \( E_{\rm sym} \) (1) remains positive and real, (2) diminishes its contribution to the overall energy, and (3) predicts \( Y_{e,\beta} \in (0, 0.5] \), with \( Y_{e,\beta} \) approaching the value from SFHo at low densities. For \( n < 0.5 \, n_s \), we adopt the following model for the symmetry energy:
\begin{equation}
\label{eq:Esymlow}
E_{\rm sym, low}(n) = [1 - \chi(n)] E_{\rm fl} + \chi(n) E_{\rm PL}(n),
\end{equation}
where \( E_{\rm PL}(n) \) represents a power-law function and \( E_{\rm fl} \) is an energy floor, set to 11.22~MeV. This floor corresponds to \( E_{\rm sym}(0.5 \, n_s)/2 \), calculated using the best-fit parameters for the SFHo EoS~\cite{Steiner:2012rk} (\( S = 31.47 \)~MeV, \( L = 47.10 \)~MeV, \( \gamma = 0.41 \); \cite{Raithel:2019gws}). In this expression, \( \chi(n) \) is a smooth transition function given by
\begin{equation}
\chi(n) = \frac{1 + \tanh\left[ X \left( n - n_0 \right) \right]}{2},
\end{equation}
where we select \( X = 40 \) and \( n_0 = 0.025 \)~fm\(^{-3}\), ensuring that \( \chi(n) \approx 1 \) at \( 0.5 \, n_s \) (to within 1\% accuracy) and \( \chi(n) \) decreases at lower densities. These parameters and the value of \( E_{\rm fl} \) were chosen empirically to ensure that \( Y_{p,\beta}(n) \) closely matches that of SFHo in this density range for the EoSs considered in this work.\\
To ensure continuity in the symmetry energy and corresponding pressure, we define the power-law energy extrapolation as
\begin{multline}
E_{\rm PL}(n) =  E_{\rm sym}(n_t) +  
\frac{P_{\rm sym}(n_t)}{n_t (\gamma_{\rm PL}-1)} \left[ \left(\frac{n}{n_t}\right)^{\gamma_{\rm PL}-1} - 1 \right],
\end{multline}
where \( n_t = 0.5 \, n_s \), and the power-law index \( \gamma_{\rm PL} \) is given by
\begin{equation}
\gamma_{\rm PL} = \frac{\partial P_{\rm sym}(n)}{\partial n} \biggr \rvert_{n_t} \left[ \frac{n_t}{P_{\rm sym}(n_t)} \right],
\end{equation} 
and \( P_{\rm sym}(n) = n^2 \frac{\partial E_{\rm sym}(n)}{\partial n} \).
For the corresponding model of the low-density symmetry pressure, we neglect the density derivatives of \( \chi(n) \), as they introduce unphysical density dependencies. Instead, we calculate the pressure in the two asymptotic limits and use \( \chi(n) \) to smoothly connect these regions:
\begin{align}
P_{\rm sym, low}(n) &= n^2 \left( \frac{\partial E_{\rm fl}}{\partial n} \right) [1 - \chi(n)] + n^2 \left( \frac{\partial E_{\rm PL}(n)}{\partial n} \right) \chi(n)  \\ \notag 
&= P_{\rm PL}(n) \chi(n),
\end{align}
where the first term vanishes since \( E_{\rm fl} \) is constant, leaving the expression:
\begin{equation}
P_{\rm PL}(n) = P_{\rm sym}(n_t) \left(\frac{n}{n_t}\right)^{\gamma_{\rm PL}}.
\end{equation}

\subsubsection{Thermal contribution to the energy} 
To describe the thermal contribution to the energy per baryon in $npe$--matter, we define, following \cite{Raithel:2019gws}, the thermal energy as a piecewise function that accounts for relativistic, ideal-fluid, and degenerate regimes. This allows capturing the dominant thermal physics across density ranges of astrophysical interest.\\
It is, therefore, convenient to express the thermal energy per particle as
\begin{equation}
\label{eq:Eth-piecewise}
\begin{aligned}
E_{\rm th}(n, Y_p, T) = 
\begin{cases}
E_{\rm rel}(n,T), & n < n_1, \\
E_{\rm ideal}(T), & n_1 < n < n_2, \\
E_{\rm th, deg}(n, Y_p=\tfrac{1}{2}, T) \\
 + E_{\rm sym, th}(n, T)(1 - 2 Y_p)^2, & n > n_2,
\end{cases}
\end{aligned}
\end{equation}
where:
\begin{align}
E_{\rm rel}(n,T) &= \frac{4 \sigma f_s T^4}{c n}, \\
E_{\rm ideal}(T) &= \frac{3}{2} k_B T.
\end{align}
Here, \( E_{\rm th, deg}(n, Y_p=1/2, T) \) is the degenerate thermal energy in symmetric matter, introduced below. Because the relativistic and ideal-fluid components do not depend on the proton fraction \( Y_p \), the symmetry-energy correction appears only in the degenerate regime.\\
The transition densities \( n_1 \) and \( n_2 \) are defined as the densities at which the relativistic and ideal-fluid thermal energies are equal, and where the ideal-fluid and degenerate thermal energies are equal, respectively. These densities depend on both temperature and composition.\\

{\em Degenerate Thermal Energy}\\
In the degenerate regime, we model nucleons as a free Fermi gas. The leading-order thermal energy per baryon for a species \( q \) is
\begin{align}
\begin{split}
\label{eq:Ethq}
E_{\rm th,~q}^{\rm deg}(n, Y_q, T) &= a(Y_q n, M^*) Y_q T^2,
\end{split}
\end{align}
where the level-density parameter is
\begin{equation}
\label{eq:a}
a(n_q, M^*) = \frac{\pi^2 k_B^2}{2} \frac{ \sqrt{ \left( 3 \pi^2 n_q \right)^{2/3} (\hbar c)^2 + M^*(n_q)^2 } }{\left( 3 \pi^2 n_q \right)^{2/3} (\hbar c)^2 },
\end{equation}
and \( M^*(n_q) \) is the Dirac effective mass. For a full derivation including next-to-leading order corrections, see \citet{Constantinou:2015zia}.\\
For symmetric matter, where \( n_n = n_p = n/2 \), the degenerate thermal energy becomes
\begin{align}
\begin{split}
E_{\rm th,~nucl}^{\rm deg}(n, T) &= a(0.5 n, M^*_{\rm SM}) T^2,
\end{split}
\end{align}
assuming \( M^*_{\rm n, SM} \approx M^*_{\rm p, SM} \approx M^*_{\rm SM} \).\\

{\em Leptonic Contribution}\\
For degenerate leptons, particularly electrons, the thermal energy is simpler. Their effective mass remains approximately constant (\( M^*_e \approx m_e \)). The degenerate electron thermal energy is
\begin{equation}
\label{eq:Ethelect}
E_{\mathrm{th,~}e^-}^{\rm deg}(n, Y_p, T) = a(Y_p n, m_e) Y_p T^2,
\end{equation}
where we impose charge neutrality (\( Y_e = Y_p \)). In the presence of positrons, this should be generalized to the net lepton fraction.\\
While the above piecewise form is physically motivated and convenient for analytic expressions, the discontinuities at the transition densities can be problematic for numerical simulations. Discontinuities may lead to unphysical reflections of matter waves at interfaces. Therefore, for practical implementations we use a smooth approximation:
\begin{align}
\label{eq:addinv}
\begin{split}
E_{\rm th}(n, Y_p, T) & \approx E_{\rm rel}(n,T) +  \\
&\left[ E_{\rm ideal}(T)^{-1} + E_{\rm th, deg}(n, Y_p, T)^{-1} \right]^{-1},
\end{split}
\end{align}
where the inverse sum ensures that \( E_{\rm ideal} \) dominates at intermediate densities and \( E_{\rm th, deg} \) dominates at high densities. This smoothed form is also computationally efficient and avoids the explicit calculation of \( n_1 \) and \( n_2 \).\\
With this, the complete energy density equation is given by
\begin{equation}
\begin{aligned}
E(n, Y_p, T) &= \underbrace{\big( \text{Cold EOS in $\beta$-equilibrium} \big)}_{\text{background}} \\
&\quad + 3K \left( Y_p^{4/3} - Y_{p,\beta}^{4/3} \right) n^{1/3} \\
&\quad + E_{\rm sym}(n, 0) \left[ (1 - 2Y_p)^2 - (1 - 2Y_{p,\beta})^2 \right] \\
&\quad + E_{\rm th}(n, Y_p, T).
\end{aligned}
\end{equation}

\subsubsection{Approximating the Dirac effective mass} With the complete set of equations for the thermal energy available, it becomes clear that calculating this thermal energy requires knowledge of the Dirac effective masses in symmetric matter. Determining these masses, in turn, depends on understanding the scalar meson interactions and particle potentials for each equation of state. Since such details go beyond the scope of creating a straightforward, physically motivated framework, an approximation for the Dirac effective mass is introduced. This approximation must be consistent with existing data, and thus, it must satisfy two limits: at low densities, the effective mass should approach the dominant nucleon mass, and at high densities, \( M^* \) should decrease due to the increased significance of particle interactions. This behavior is captured using the following power-law expression:
\begin{equation}
\label{eq:Meff}
M^*(n_q) = \left\{ (m c^2)^{-b} + \left[ mc^2 \left( \frac{ n_q }{n_0 }\right)^{-\alpha} \right]^{-b} \right\}^{-1/b},
\end{equation}
where \( m \) represents the dominant nucleon mass (taken as the neutron mass \( m c^2 = 939.57 \) MeV), and \( n_0 \) is the density above which \( M^* \) begins to decrease. The exponent \( b \), which controls the steepness of the transition to the region where the effective mass decreases, is fixed at \( b = 2 \). Lastly, \( \alpha \) is the parameter that defines the slope at high densities. With \( b \) fixed, the effective mass is characterized by two free parameters: \( n_0 \) and \( \alpha \).

\subsubsection{The complete pressure equations}
Using the energy relations from the previous subsection, the expression for the pressure can be obtained. The pressure for $n$-$p$-$e$ matter is given by 
\begin{align}
    P(n, Y_p, T) &= P_{\text{cold}}(n,Y_{p,\beta}) + K \left( Y_p^{4/3} - Y_{p,\beta}^{4/3} \right) n^{4/3} \notag \\
    &\quad + P_{\text{sym}} (n,T=0) \left[ (1-2Y_p)^2 - (1-2Y_{p,\beta})^2 \right] \notag \\
    &\quad + P_{\text{th}}(n, Y_p, T).
\end{align}
where the thermal pressure contribution is given by  
\begin{align}
P_{\text{th}}(n, Y_p, T) &=
\begin{cases}  
\frac{4 \sigma_{\text{fs}} T^4}{3c}, & n < n_1, \\  
nk_B T, & n_1 < n < n_2, \\
-\frac{\partial a(0.5 n, M_{\text{SM}})}{\partial n} \\
+ \frac{\partial a(Y_p n, m_e)}{\partial n} Y_p n^2 T^2, & n > n_2.
\end{cases}  
\end{align}
Here, $n_1$ and $n_2$ are the transition densities for thermal pressure at a given temperature and proton fraction. \\ 
The symmetry pressure is expressed as  
\begin{align}
\label{eq:Psym}
P_{\rm sym}(n, T=0) &= \frac{2\eta}{3} n E_{\rm sym}^{\rm kin}(n) \\
& + \gamma n \left[ S_0 - \eta E_{\rm sym}^{\rm kin}(n_{\rm sat}) \right] \left( \frac{n}{n_{\rm sat}} \right)^{\gamma} .
\end{align}
As before, we avoid the piecewise expressions for the thermal pressure by using the smooth approximation
\begin{align}
P_{\text{th}}(n,Y_p,T) &\approx P_{\text{rel}} + \left( P_{\text{ideal}}^{-1} + P_{\text{deg}}^{-1} \right)^{-1}.
\end{align}
%%%%%%%%%%%%%%%%%%%%%%%%%%%%%%%%%%%%
\begin{figure*}[ht!]
    \centering
    \includegraphics[width=\linewidth]{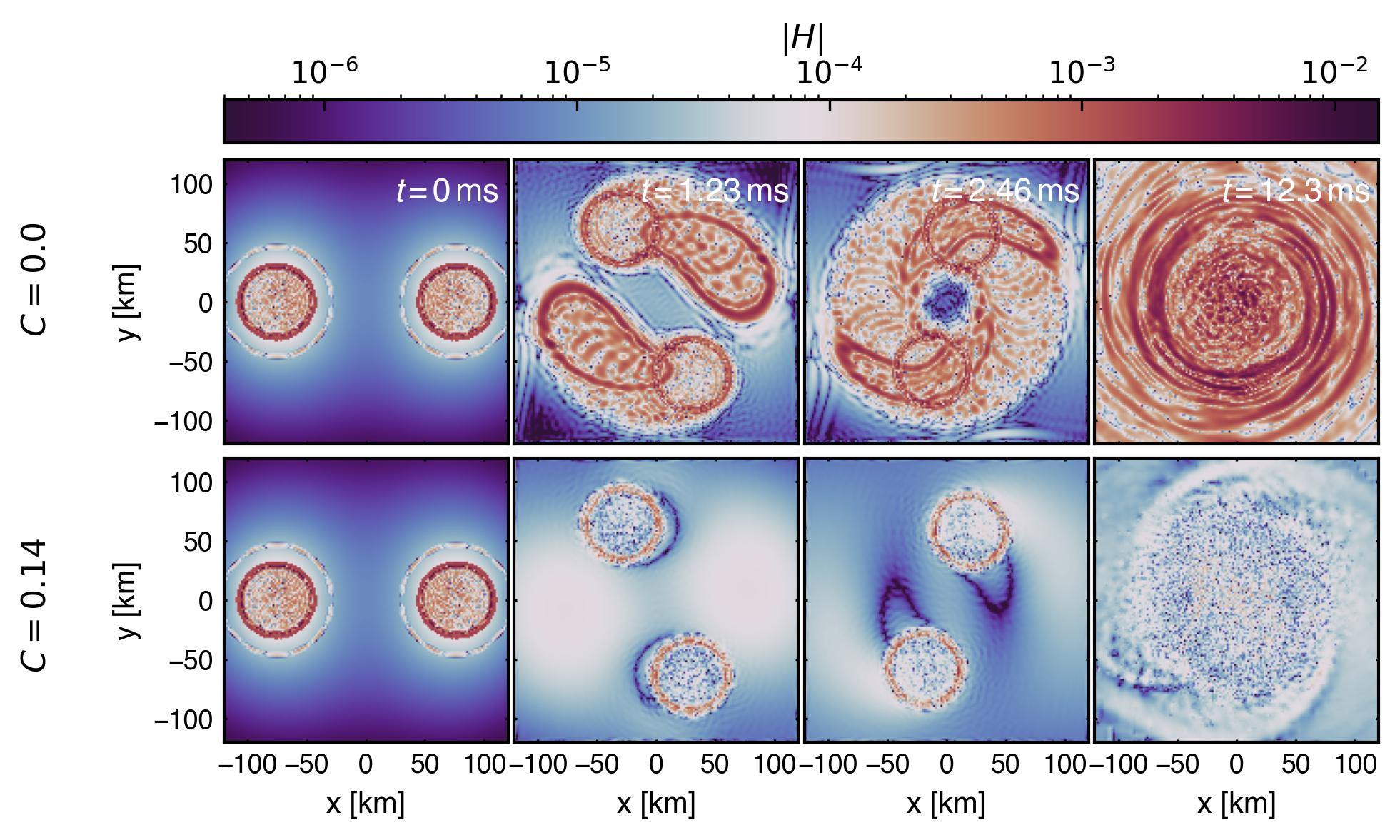}
    \caption{A comparison of the absolute value of the Hamiltonian
    constraint in the xy-plane at different times for the 1 million
    particle runs with no constraint damping (row 1) with the
    corresponding run with constraint damping (row 2). The first column is
    for the initial time $T=0$ ms, the second column is for $T=1.23$ ms, the
    third column is for $T=2.46$ ms and the fourth column is for $T=12.3$ ms.}
    \label{fig:ham_2d}
\end{figure*}
%%%%%%%%%%%%%%%%%%%%%%%%%%%%%%%%%%%
%
%%%%%%%%%%%%%%%%%%%%%%%%%%%%%%%%%%%
\begin{figure}[t]
  \centering
  \includegraphics[width=\linewidth]{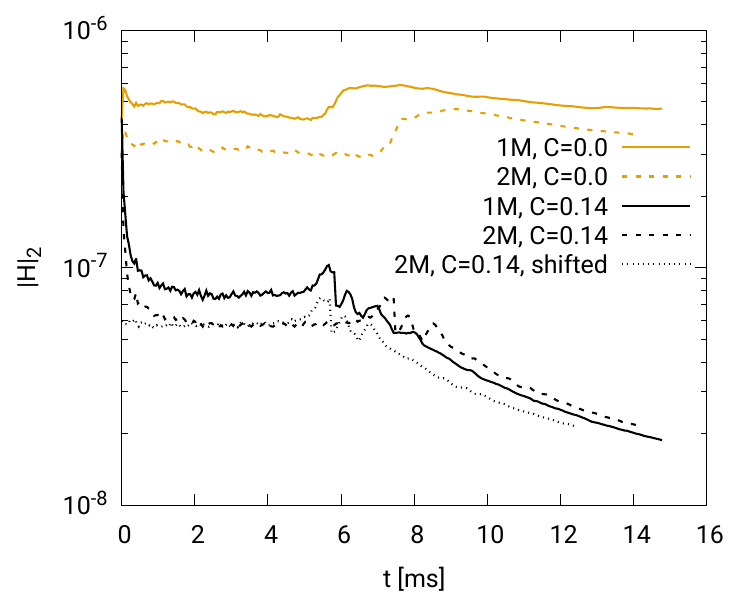}
  \caption{The $L^2$-norm of the Hamiltonian constraint as function of
  time. In this plot no constraint damping is shown with the orange 
  curves, while full constraint damping is shown with the black curves.
  Runs with one million particles are shown with solid curves, while two
  million particles are shown with dashed curves. Finally the dotted line
  shows the two million particle run with constraint damping shifted in time
  so that the merger time is aligned with the one-million-particle run.}
  \label{fig:ham_norm}
\end{figure}
%%%%%%%%%%%%%%%%%%%%%%%%%%%%%%%%%%%

\subsubsection{The complete entropy equations}
The \textit{entropy density} \( s(n, n_p, n_n, n_e, T) \), accounting for relativistic, ideal-fluid, and degenerate regimes, is given by
\begin{align}
\begin{split}
s(n,n_p, n_n,n_e, T) =
 \begin{cases}
s_{\rm rel},    &\phantom{n_1<} n < n_1 \\
s_{\rm ideal},  &n_1< n <n_2 \\
s_{\rm deg},    &\phantom{n_1<} n > n_2,
\end{cases}
\end{split}
\end{align}
where \( n_1 \) and \( n_2 \) are the transition densities for thermal energy.\\
The entropy density of a gas of relativistic leptons and photons is
\begin{equation}
s_{\rm rel} = \frac{16 \sigma f_s}{3 c} (n_p + n_n) T^3.
\end{equation}
The entropy density of a monatomic ideal fluid is given by the Sackur-Tetrode equation:
\begin{multline}
s_{\rm ideal} =  \left( n_p + n_n + n_e \right) k_B \\
\times \left\{ \ln\left[ \left( \frac{n_p + n_n}{n_p + n_n + n_e} \right) n^{-1} \left( \frac{m k_B T}{2\pi \hbar^2} \right)^{3/2} \right] + \frac{5}{2} \right\}.
\end{multline}
In the degenerate regime, the entropy density of a Fermi gas of particle species \( q \) is
\begin{equation}
s_q = 2 a_q n_q T,
\end{equation}
so that the total degenerate entropy density becomes
\begin{equation}
s_{\rm deg} = 2 \left\{ a(0.5 n, M^*_{\rm SM}) (n_p + n_n) + a(Y_p n, m_e) n_e \right\} T.
\end{equation}
To provide a continuous interpolation across the relativistic, ideal, and degenerate regimes, we adopt, once more, a smooth approximation 
\begin{align}
s(n, Y_p, T) &\approx s_{\text{rel}} + \left( s_{\text{ideal}}^{-1} + s_{\text{deg}}^{-1} \right)^{-1}.
\end{align}

\subsubsection{Numerical implementation}
In Appendix A of \cite{rosswog22b} we outlined in 
detail how we recover the physical variables from 
the conserved variables for the case where the 
cold part of the EOS is given by a piecewise–polytropic 
representation \cite{Read:2008iy} and the thermal 
part is given by an ideal gas-type law with a fixed 
thermal exponent $\Gamma_{\rm th}$. Starting from this algorithm, we change the thermal parts to the
finite–temperature extension described above.
 Starting from an initial guess for the pressure, we compute the specific internal energy, separate the cold and thermal components, determine the temperature from the thermal model, and update the thermal pressure accordingly. The pressure is then corrected until the cold and thermal contributions self-consistently match the evolved state. Both the pressure and temperature inversions are solved using \emph{Ridders’ root–finding method}~\cite{ridders1982accurate,press1992numerical},
 which is a powerful variant of the {\em regula falsi}.
 This method has a number of desirable properties, for example, it does not need derivatives, it does not jump
 out of the brackets, is generally very robust and it converges fast.
%%%%%%%%%%%%%%%%%%%%%%

\section{Results}
\label{sec:results}

We demonstrate here the new elements in \spB. In all cases we use the ENG EOS \cite{engvik96}
which is consistent with known constraints on the neutron star equation of state \cite{Biswas:2021pvm}.
Each simulation is performed for an equal-mass binary with component masses 
of either $1.3\,M_\odot$ or $1.4\,M_\odot$, starting from quasi-circular orbits at an initial coordinate 
separation of $45$~km, where the initial data were constructed using the FUKA
\cite{papenfort21} library and the code \texttt{SPHINCS\_ID} \cite{diener22a,rosswog23a} was used for the particle placement.\\
We first demonstrate the effect of the constraint damping terms, see Sec.~\ref{sec:eff_of_constraint_damping}, then
we discuss thermal effects in a merger, see Sec.~\ref{sec:thermal_effects}, and whether it is likely to determine
the involved parameters from potential
post-merger observations with future
GW-facilities. We finally summarize our results in Sec. \ref{sec:summary}.

%%%%%%%%%%%%%%%%%%%%%%%%%%%%%%%%%%%
\begin{figure*}[ht]
  \hspace*{-0.8cm}\includegraphics[width=0.5\linewidth]{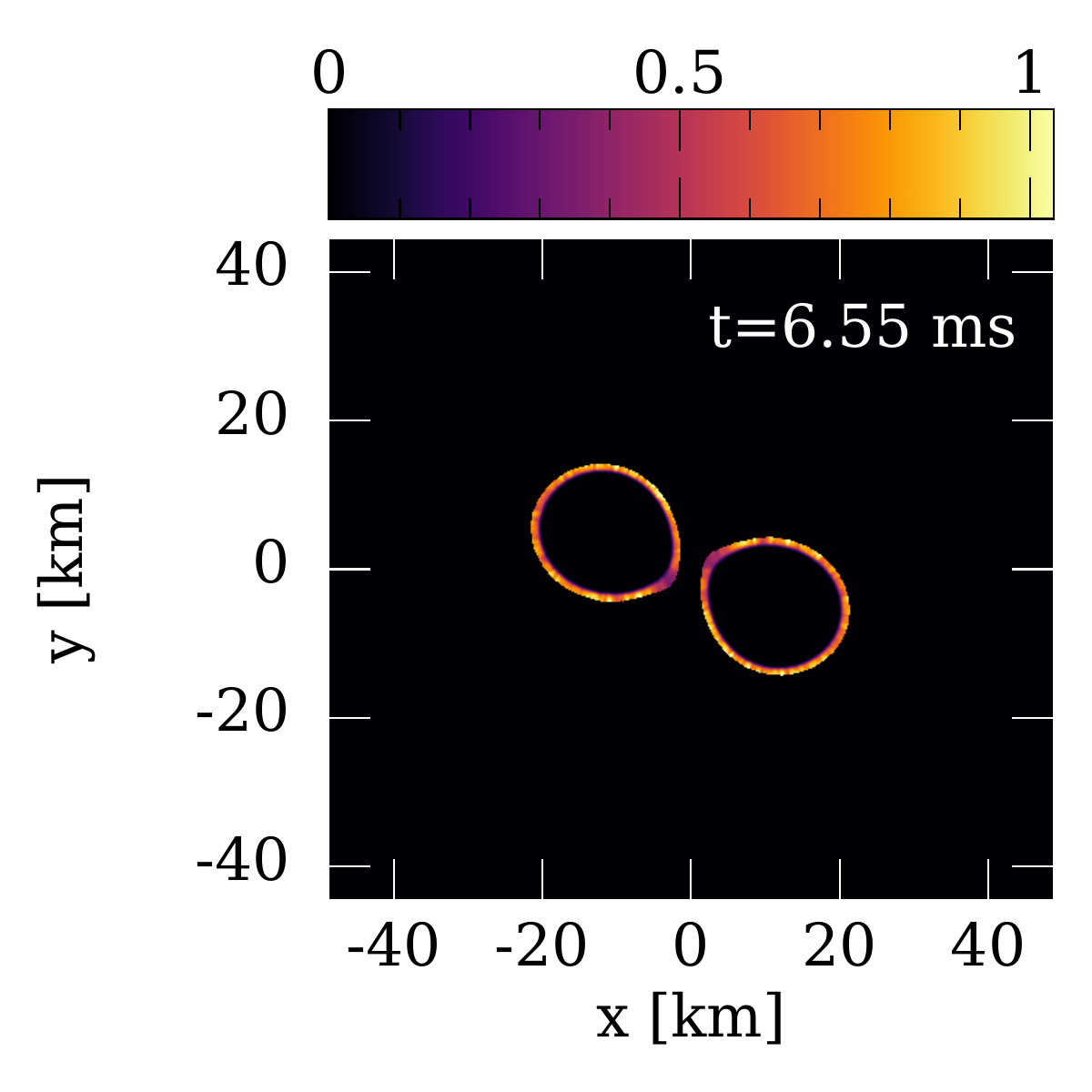}
 \hspace*{-0.4cm}\includegraphics[width=0.5\linewidth]{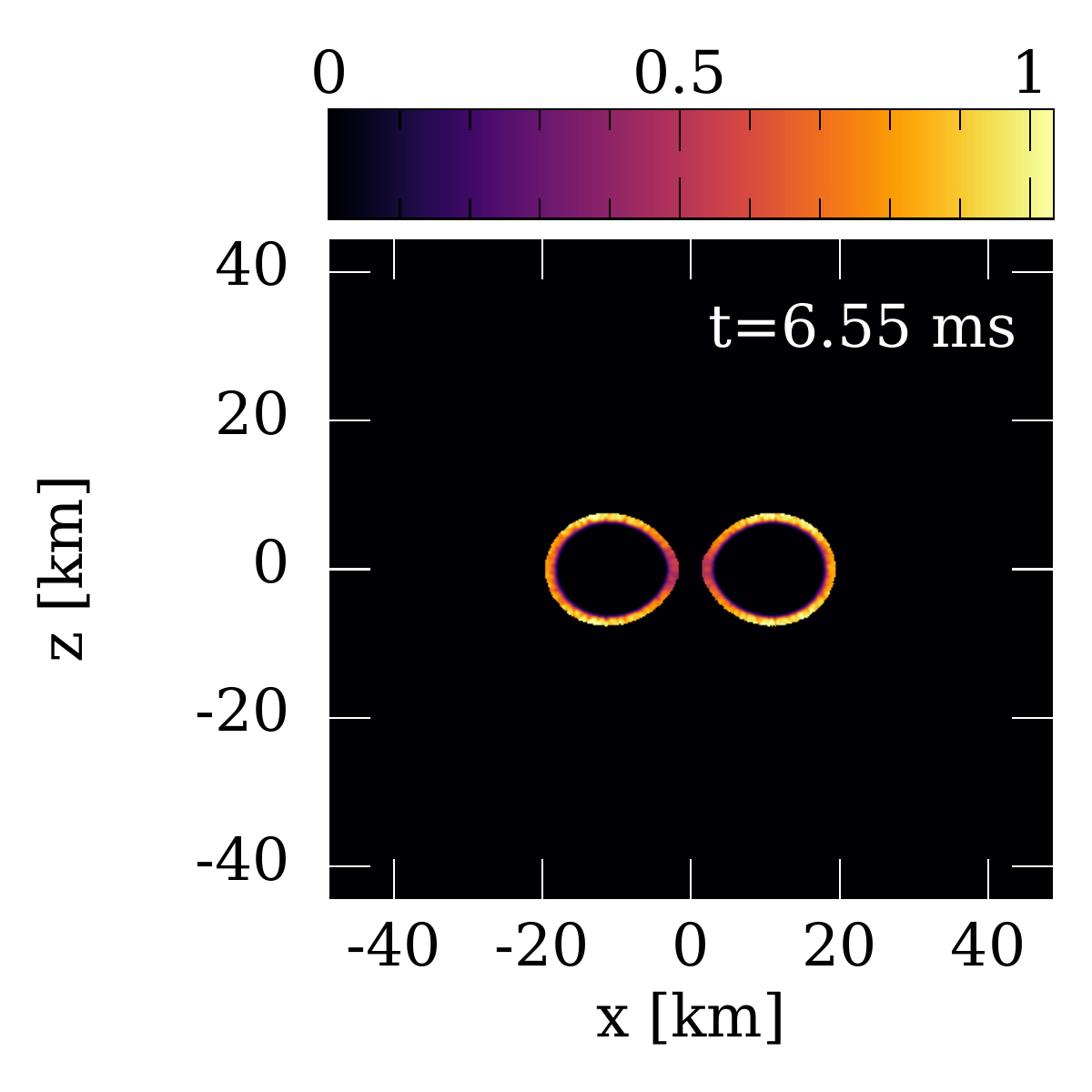}
  \caption{Temperature  distribution (in MeV) the illustrative example (2 $\times$ 1.4 \msun, three million particles; see text for more details) just before the merger.}
  \label{fig:Temp_premerger}
\end{figure*}
%%%%%%%%%%%%%%%%%%%%%%%%%%%%%%%%%%%
%
%%%%%%%%%%%%%%%%%%%%%%%%%%%%%%%%%%%
\begin{figure*}[ht]
  \centering
  \hspace*{-1.2cm}\includegraphics[width=1.1\linewidth]{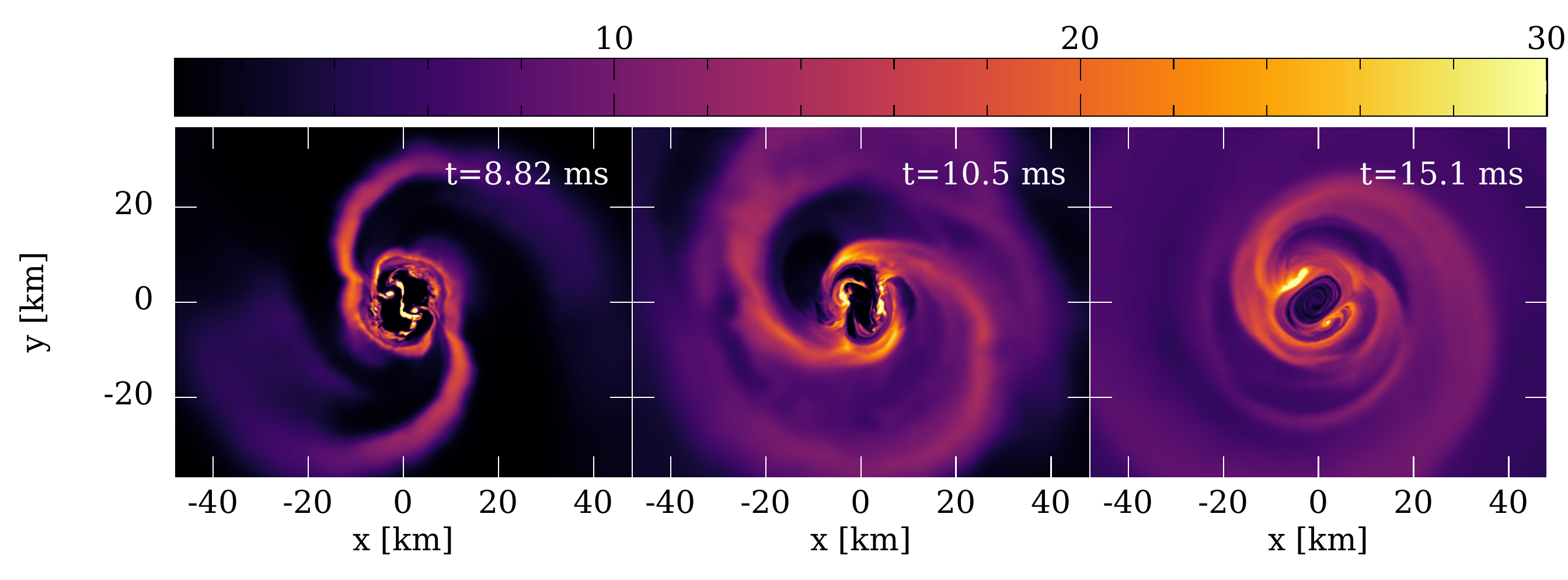}
\hspace*{-1.2cm}\includegraphics[width=1.1\linewidth]{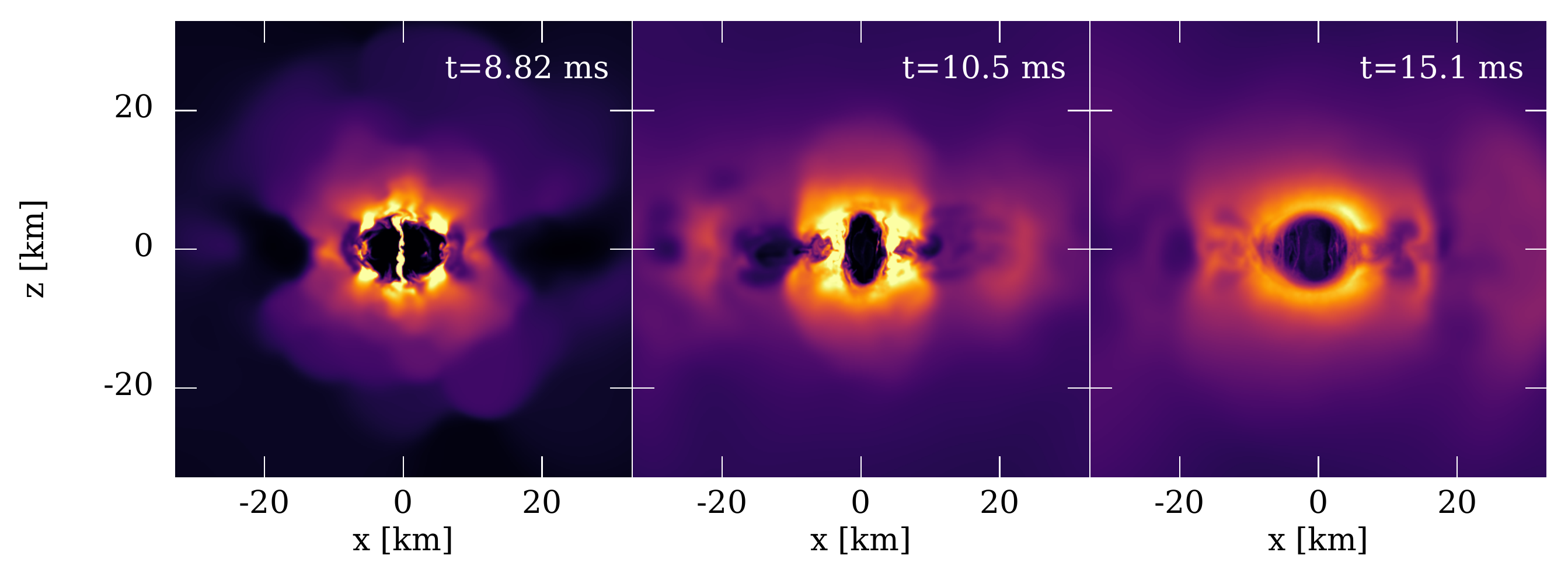}
  \caption{Temperature distribution (first row: $XY$-plane, second row:$XZ$-plane) in a merger of $2 \times 1.4$ \msun. For visibility reasons the color bar is restricted to values below 30 MeV. Very locally and for a short time (around the time shown in the first panel), peak temperatures exceed 80 MeV.}
  \label{fig:Temp_illustration}
\end{figure*}
%%%%%%%%%%%%%%%%%%%%%%%%%%%%%%%%%%%
%
\subsection{Effect of Constraint Damping}
\label{sec:eff_of_constraint_damping}
To illustrate the effect of the constraint damping  we performed simulations with 
one and two million SPH particles for a baseline setup ($2 \times 1.3$ \msun, ENG equation of state),  both without 
($C=0$) and with ($C=0.14$) constraint damping turned on, where $C$ refers 
to the strength of the constraint damping as defined in Eq.~\eqref{eq:cdamp}.\\
In Fig.~\ref{fig:ham_2d}, we show a comparison of the absolute value of the
Hamiltonian constraint violation in the orbital plane for runs without
constraint damping (row 1) and with constraint damping 
(row 2) at times $t=0$ ms (column 1), $t=1.23$ ms (column 2),
$t=2.46$ ms (column 3) and $t=12.3$ ms (column 4). For both runs, it is clear
that the constraint violations are mostly triggered by the sharp surfaces of the
stars. The main difference is that in the undamped case the constraint 
violations persist, due to the zero speed constraint mode in the BSSN 
equations, when the stars move away, leaving a significant trail of 
constraint violations, whereas the constraint violations are clearly 
damped behind the stars when the  damping
is turned on. So, without constraint damping, the whole interior of the 
orbit is eventually filled with constraint violations. After the merger,
there is no longer a sharp surface, as low density material 
surrounds the merger remnant. At late times after the merger, the 
constraint violations are significantly reduced in the constraint damped
case compared to the undamped case.
Since we are using a global timestep, we can increase the strength of the
constraint damping on the coarser grids (for every factor of 2 in 
resolution, we can increase the strength by a factor of 4). 

%%%%%%%%%%%%%%%%%%%%%%%%%%%%%%%%%%%
\begin{figure*}[t]
    \centering
    \includegraphics[width=0.95\textwidth]{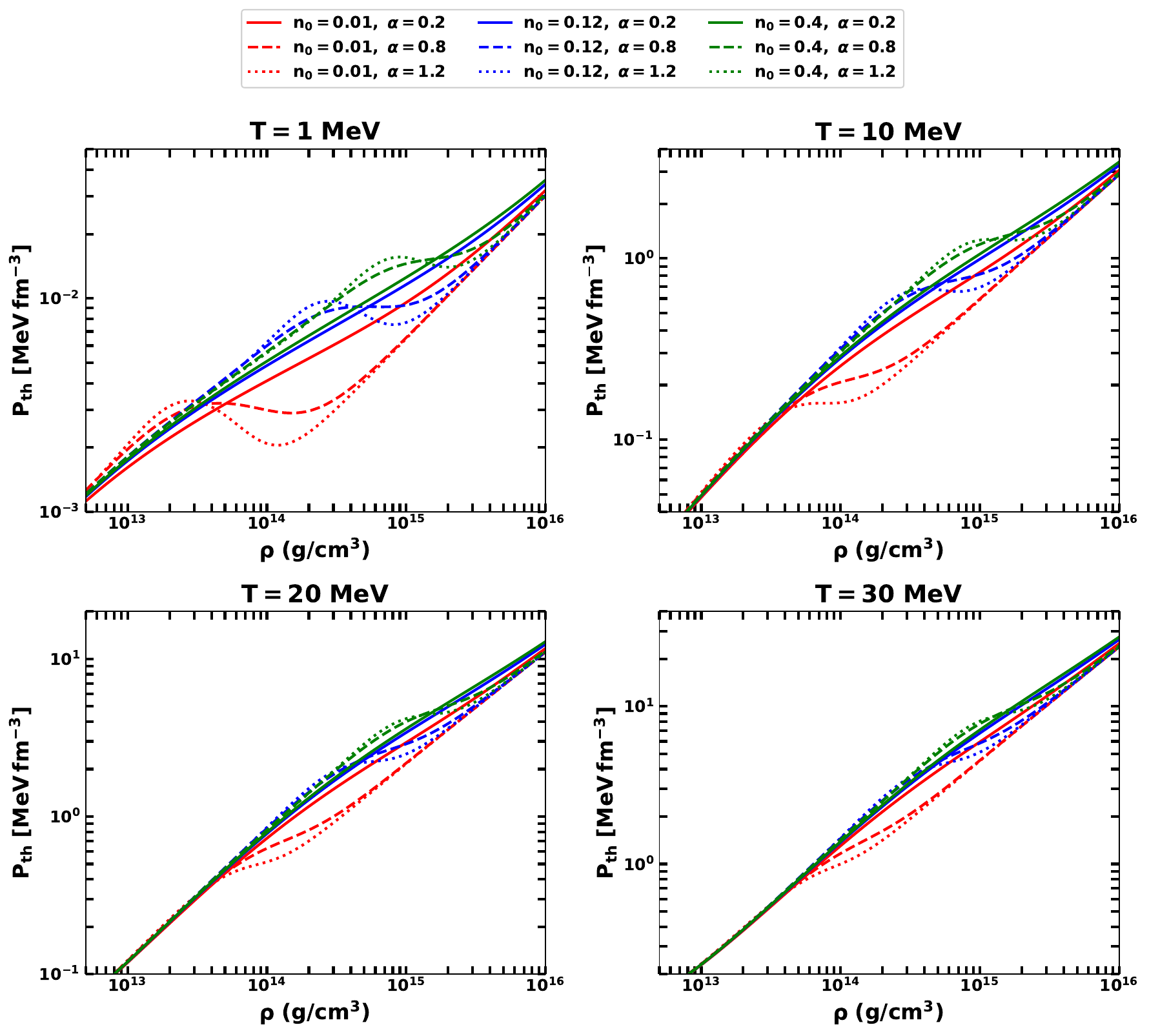}
    \caption{
        Thermal pressure $P_{\rm th}$ as a function of density $\rho$ for four temperatures
        ($T = 1, 10, 20, 30$ MeV) at $Y_e=0.1$. Note that for visibility reasons the y-axis ranges are different.
    }
    \label{fig:thermal-pressure-grid}
\end{figure*}
%%%%%%%%%%%%%%%%%%%%%%%%%%%%%%%%%%%
%
%%%%%%%%%%%%%%%%%%%%%%%%%%%%%%%%%%%
\begin{figure}[ht]
    \centering
    \includegraphics[width=\linewidth]{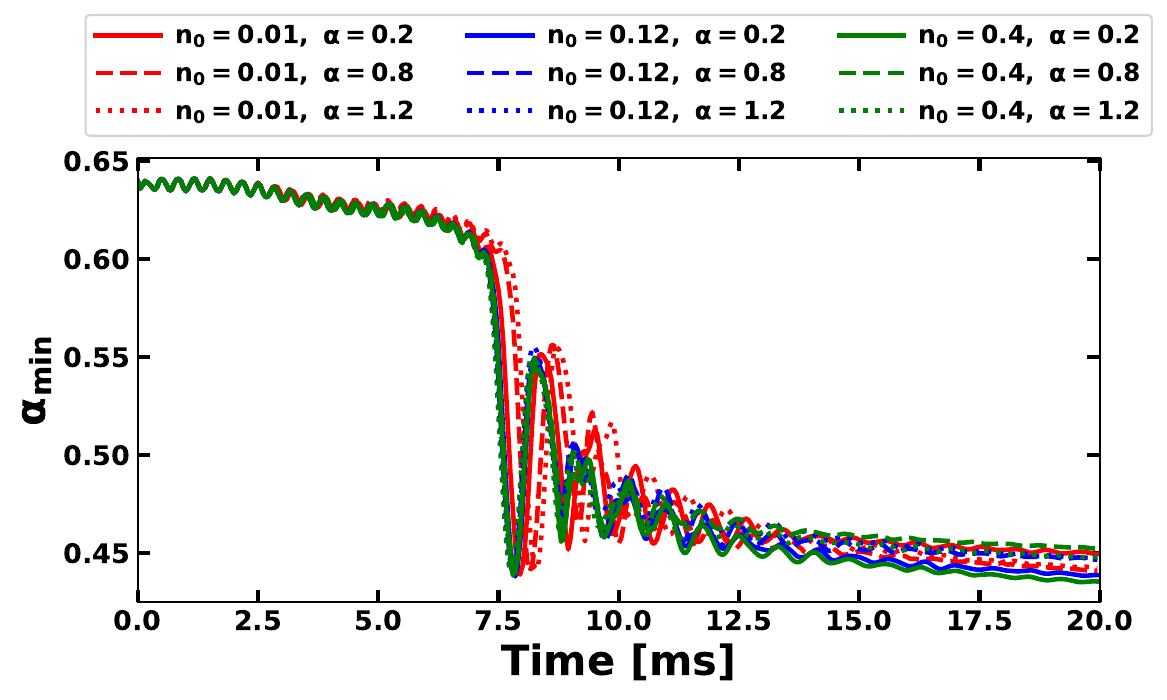}
    \caption{Time evolution of the minimum lapse $\alpha_{\rm min}(t)$
    for different thermal EOS parameterizations $(n_0,\alpha)$ at
    $2\times10^6$ particles. The lapse remains stable at late times,
    indicating that the remnant does not collapse within the simulated
    duration for any of the thermal treatments considered here.}
    \label{fig:alphamin}
\end{figure}
%%%%%%%%%%%%%%%%%%%%%%%%%%%%%%%%%%%
%
%%%%%%%%%%%%%%%%%%%%%%%%%%%%%%%%%%%
\begin{figure}[ht]
    \centering
    \includegraphics[width=\linewidth]{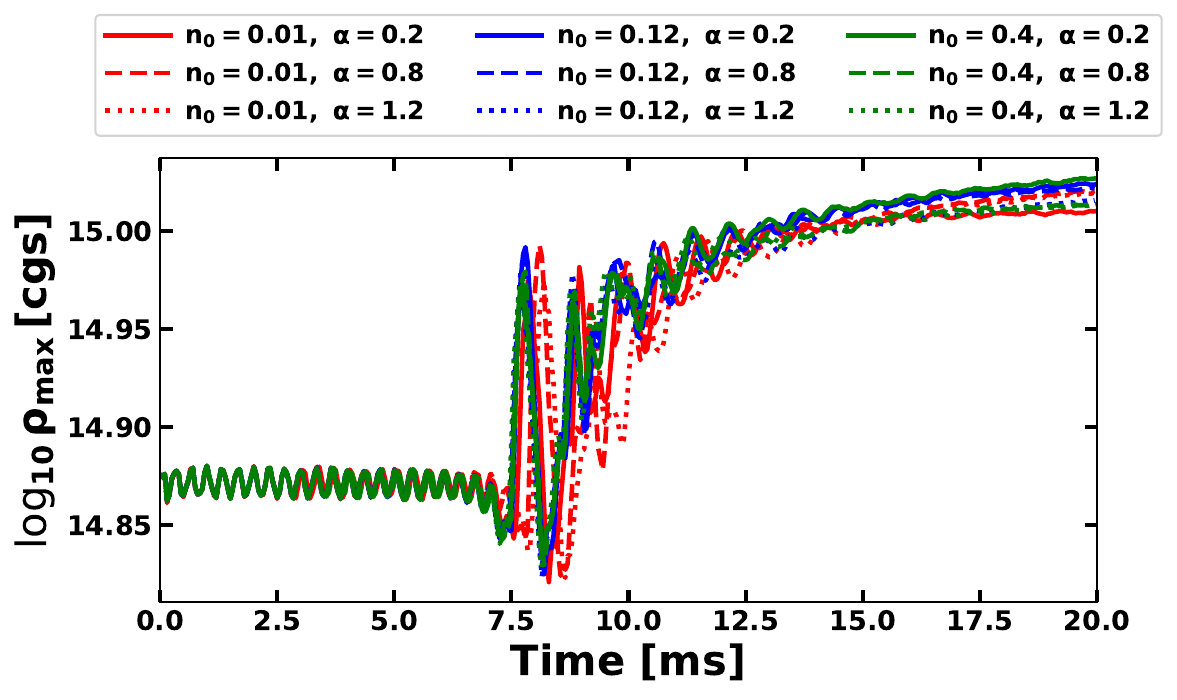}
    \caption{Time evolution of the maximum rest-mass density
    $\rho_{\rm max}(t)$ for different thermal EOS parameterizations
    $(n_0,\alpha)$ at $2\times10^6$ particles. Oscillations are clearly
    seen in the first few milliseconds after merger, but $\rho_{\rm max}$
    remains bounded at late times, confirming that no collapse occurs
    within the simulation window.}
    \label{fig:rhomax}
\end{figure}
%%%%%%%%%%%%%%%%%%%%%%%%%%%%%%%%%%%%%%
\begin{figure}[ht]
    \centering
    \includegraphics[width=0.97\linewidth]{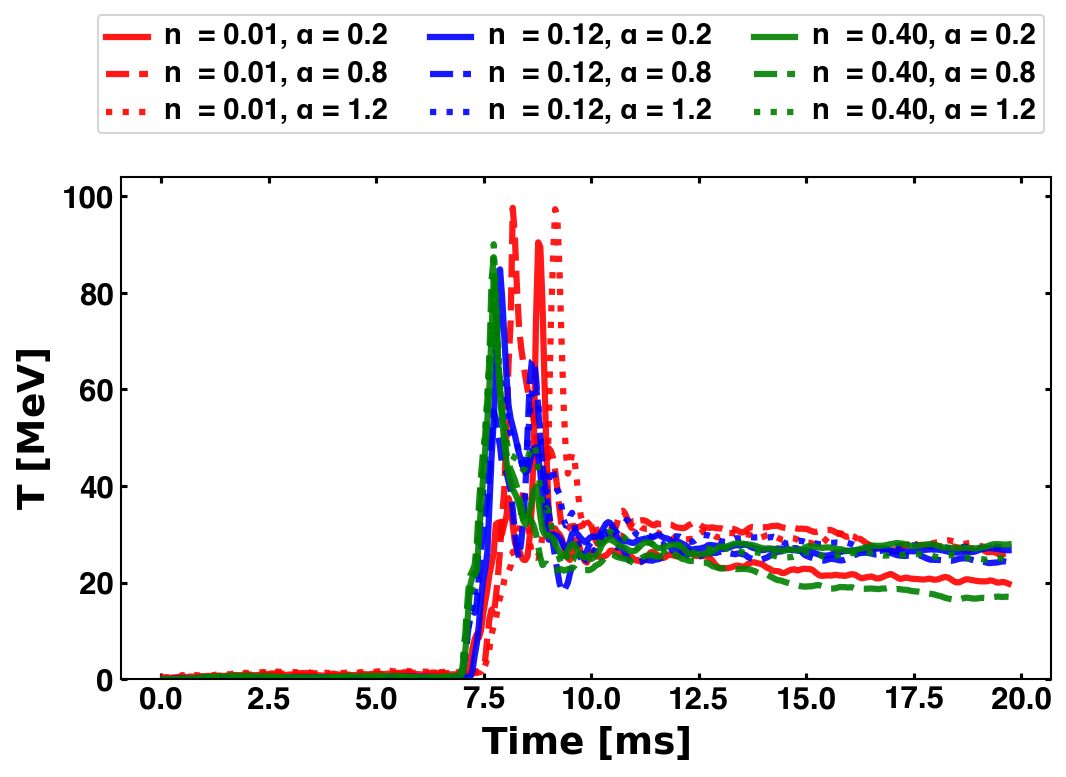}
    \caption{Average temperature of the 100 hottest SPH particles as a
    function of time for the explored thermal EOS parameterizations.}
    \label{fig:Thot100}
\end{figure}
%%%%%%%%%%%%%%%%%%%%%%%%%%%%%%%%%%
\begin{figure}[ht]
    \centering
    \includegraphics[width=0.97\linewidth]{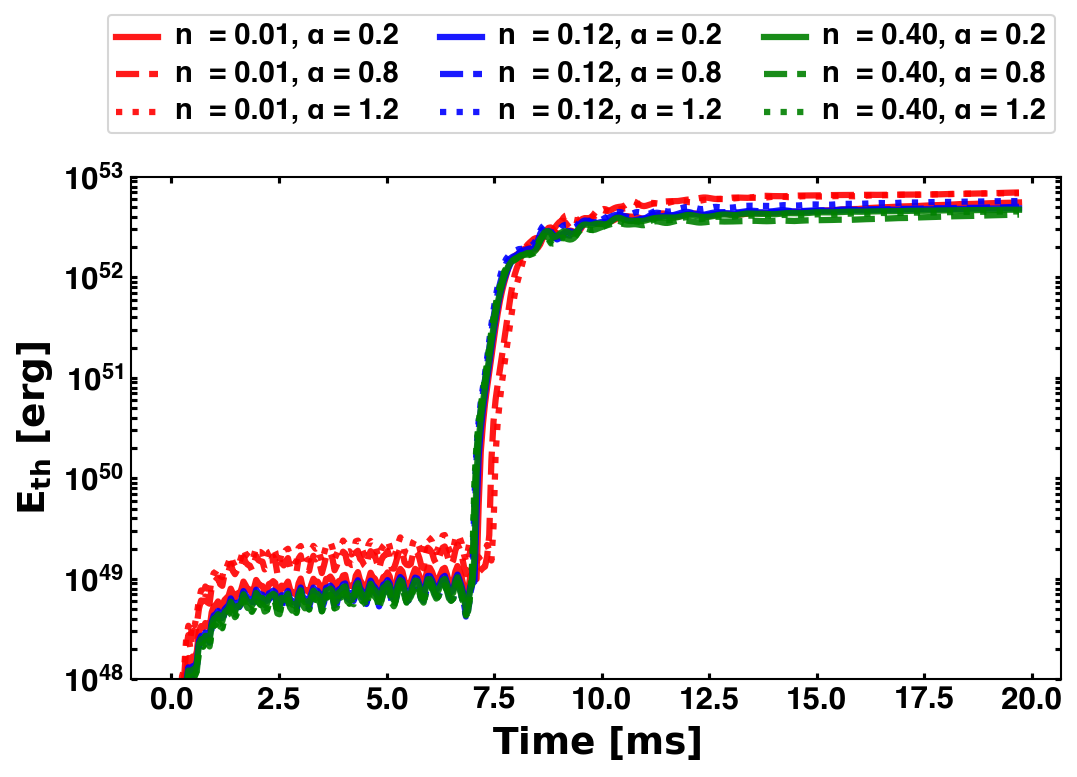}
    \caption{Integrated thermal energy in the high-density region
    $\rho>10^{14}\ \mathrm{g\ cm^{-3}}$ as a function of time for different
    $(n_0,\alpha)$ choices.}
    \label{fig:EthHigh}
\end{figure}
%%%%%%%%%%%%%%%%%%%%%%%%%%%%%%%%%%%
%
%%%%%%%%%%%%%%%%%%%%%%%%%%%%%%%%%%
\begin{figure}[t]
    \centering
    \includegraphics[width=\linewidth]{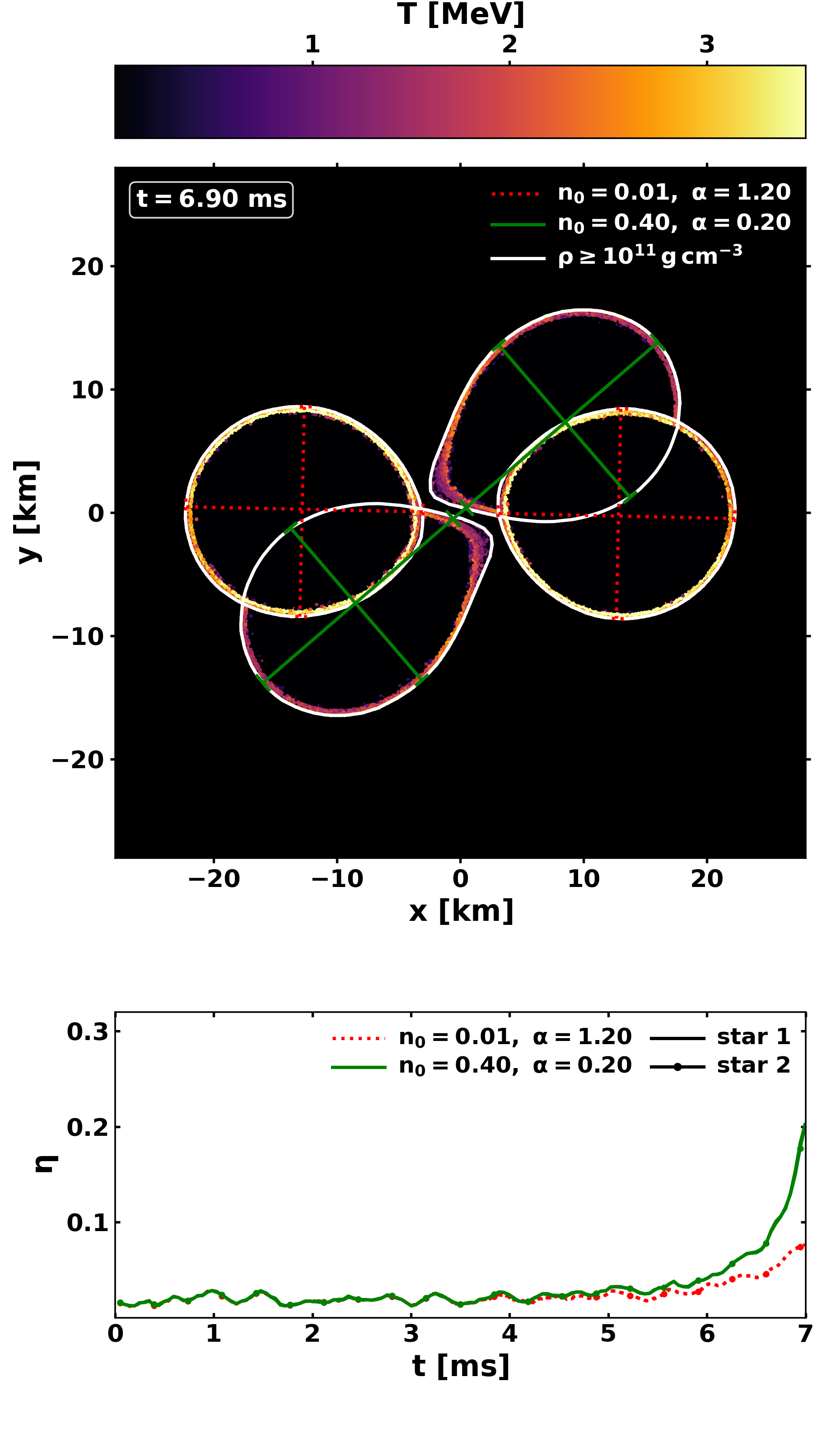}
    \caption{
    Tidal deformation in the two simulations with two million particles and the earliest and latest merger times,
    $(n_0,\alpha)=(0.4,0.2)$ and $(0.01,1.2)$.
    \textit{Top:} orbital-plane snapshot at $t= 6.90$~ms. The
    $\rho=10^{11}\,{\rm g\,cm^{-3}}$ contour outlines the high-density bulk of each NS.
    The in-plane axes indicate the line connecting the COMs and the
    perpendicular direction within the orbital plane.
    \textit{Bottom:} inspiral evolution of the deformation measure $\eta(t)$ of both merging stars for the same
    two runs.}
    \label{fig:tidal_elongation}
\end{figure}
%%%%%%%%%%%%%%%%%%%%%%%%%%%%%%%%%%
%
%%%%%%%%%%%%%%%%%%%%%%%%%%%%%%%%%%%
\begin{figure}[t]
    \centering
    \includegraphics[width=\linewidth]{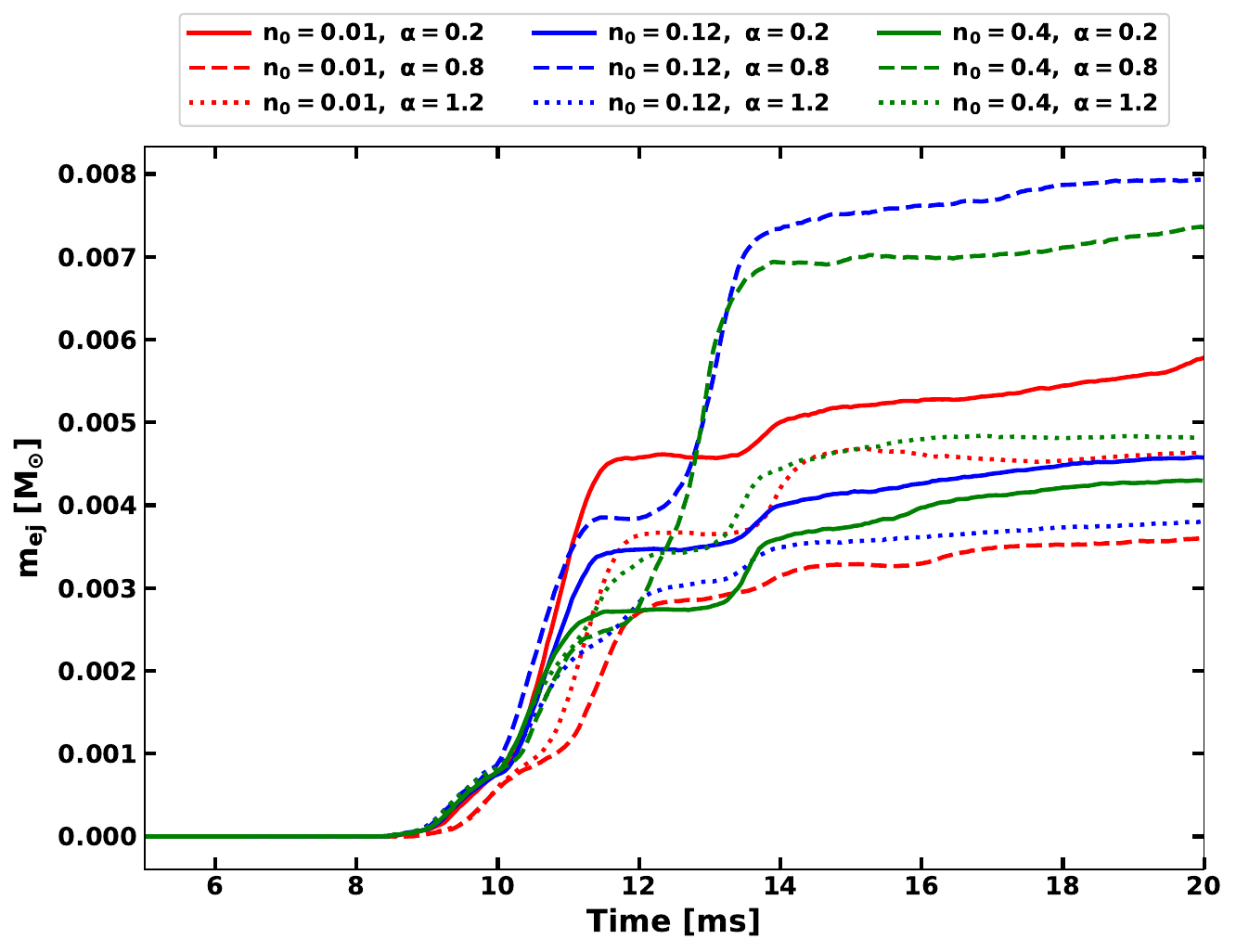}
    \caption{Time evolution of the dynamical ejecta mass for different
    thermal EOS parameterizations $(n_0,\alpha)$ at $2\times10^6$
    particles.}
    \label{fig:mej}
\end{figure}
%%%%%%%%%%%%%%%%%%%%%%%%%%%%%%%%%%%
%
%%%%%%%%%%%%%%%%%%%%%%%%%%%%%%%%%%%
\begin{figure*}[ht]
    \centering
    \includegraphics[width=0.97\linewidth]{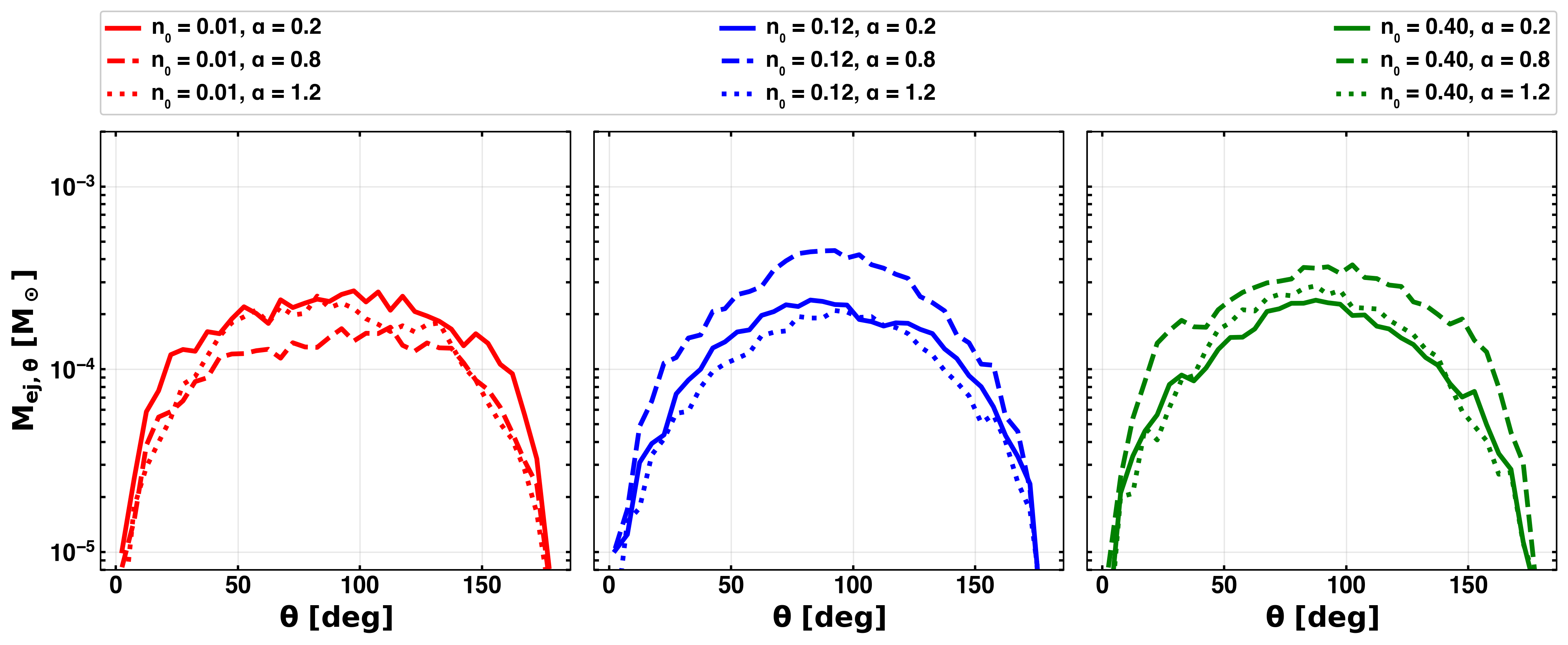}

    \vspace{0.3cm}

    \includegraphics[width=0.97\linewidth]{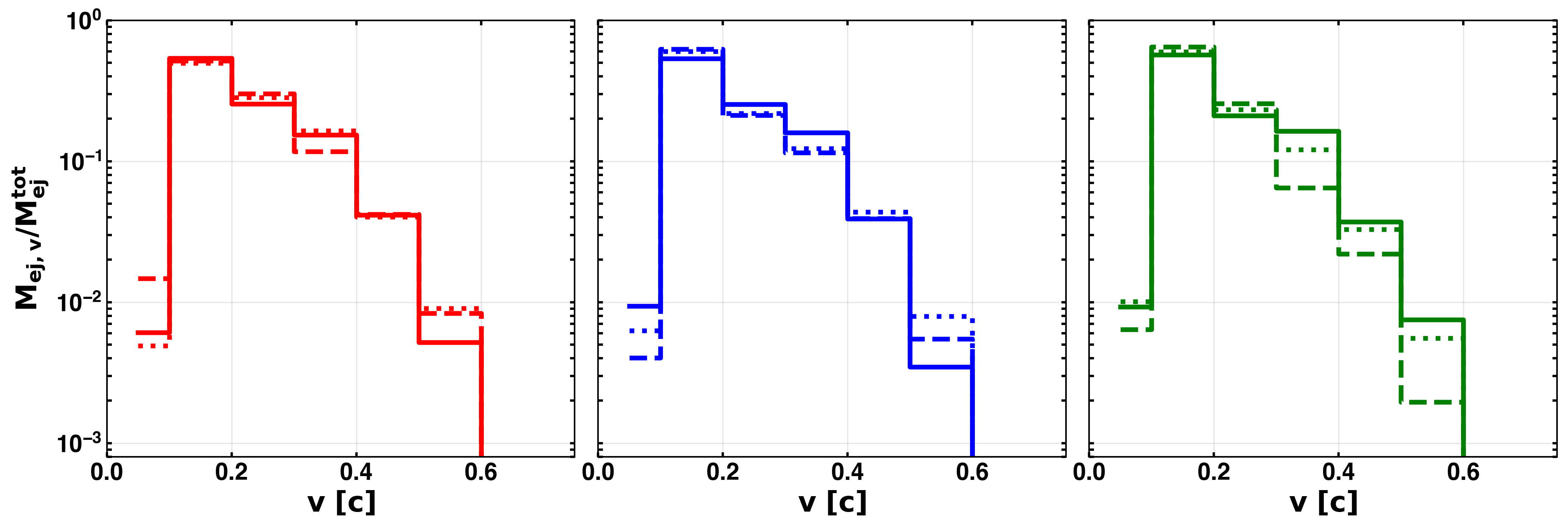}

    \caption{Angular (top) and velocity-bin (bottom) ejecta profiles for different $(n_0,\alpha)$.
    Both panels show only minor variations across the parameter space.}
    \label{fig:ejecta_profiles}
\end{figure*}
%%%%%%%%%%%%%%%%%%%%%%%%%%%%%%%%%%%
%
%%%%%%%%%%%%%%%%%%%%%%%%%%%%%%%%%%%
\begin{figure}[t]
    \centering
    \includegraphics[width=\linewidth]{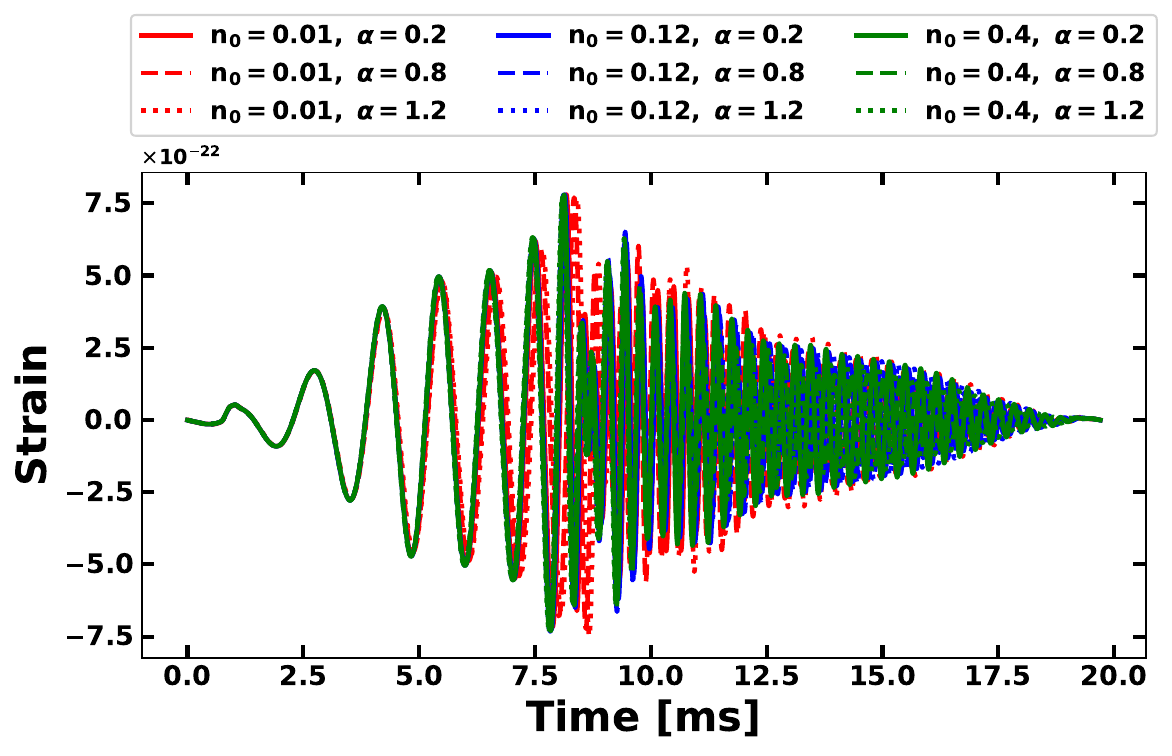}
    \caption{Plus polarization of the $(2,2)$ mode of the gravitational-wave
    strain from simulations with $2\times10^6$ particles for different
    thermal EOS parameterizations $(n_0,\alpha)$. The inspiral portions
    overlap nearly perfectly, while post-merger oscillations show modest
    but systematic differences in phase evolution and damping.}
    \label{fig:hplus}
\end{figure}
\begin{figure}[t]
    \centering
    \includegraphics[width=\linewidth]{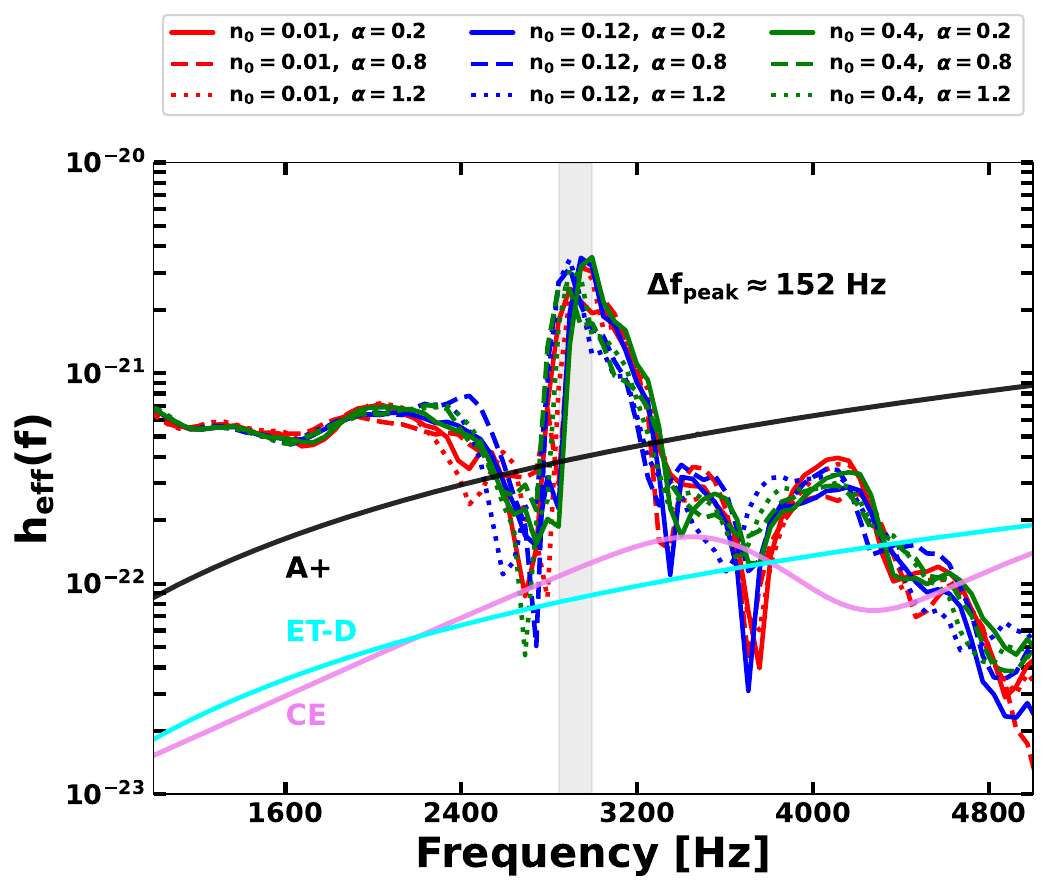}
    \caption{Effective amplitude spectra corresponding to the waveforms in
    Fig.~\ref{fig:hplus}, compared with Advanced LIGO A+, Einstein telescope (ET), and Cosmic
    Explorer (CE) design sensitivities (40 Mpc). All models display a
    dominant kHz peak associated with the fundamental quadrupolar mode of
    the remnant. Thermal EOS variations induce few-percent shifts in the
    peak frequency and alter the prominence of secondary features.}
    \label{fig:psd}
\end{figure}
%%%%%%%%%%%%%%%%%%%%%%%%%%%%%%%%%%%
%
In Fig.~\ref{fig:ham_norm} we
show the $L^2$ norm of the Hamiltonian constraint as a function of time.
In this plot the orange curves are for runs without constraint damping 
($C=0$) while black curves are for runs with constraint damping 
($C=0.14$), solid curves are for runs with one million particles, dashed 
curves are for two million particles, and the dotted line is for the 
constraint damped run with two million particles shifted in time to align
the merger time with the one million particle run.

In the undamped case, the $L^2$ norm of the constraint violations grows
rapidly initially, then stays roughly constant during the inspiral phase,
grows again during the merger, and then only decreases slightly after the 
merger. In the damped case, on the other hand, the constraint violations
drop rapidly from the initial value, then stay roughly constant during
the inspiral (at a level about 5 to 6 times lower than the undamped case),
increase slightly during the merger, but with the disappearance of the 
sharp neutron star surfaces, continues to decrease significantly, so that at the
end they are about 20 times smaller than in the undamped case. The dotted 
curve (shifting the two million particle constraint damped case in time to 
align the merger with the one million particle case), have been added to 
make it clear that the constraint violations decrease with resolution. 

\subsection{Effect of Thermal Equation of State}
\label{sec:thermal_effects}
We will first discuss the temperature evolution of generic neutron star 
merger (2$\times$1.4 \msun, no initial spins, ENG equation of state),
and then we systematically explore the 
impact of the Dirac effective mass parameters $n_0$ and $\alpha$ in 
the thermal EOS. We also
monitor the impact on ejecta and we  explore 
the imprint on the resulting post-merger GW-signal.

\subsubsection{Illustrative case}
We first focus on a generic neutron star merger of two 1.4 \Msun neutron stars with the ENG-EOS
and a fiducial set of parameters for the Dirac mass $M^\ast$, $n_0=0.1$, 
$\alpha=0.9$ in  Eq.~(\ref{eq:Meff}). During the inspiral the temperatures remain 
at essentially zero. The particles only
reach temperatures of $\sim$ 1 MeV in the surface 
layers, within the outermost few hundred meters, where the 
densities drop by orders of magnitude and the particles 
have to continuously adjust to the dynamically  changing tides, see Fig.~\ref{fig:Temp_premerger}. This temperature is about an order on magnitude
lower than what is typically found in Eulerian numerical relativity simulations,
see e.g. \citep{hammond21,gittins25}.
During
the merger the shear layer between the stars is Kelvin-Helmholtz unstable, and peak temperatures  
exceeding 80 MeV are reached (very locally) in the resulting vortices during the maximum compression phase, 
approximately corresponding to the snapshot shown in the first column
in Fig.~\ref{fig:Temp_illustration}. The central regions of the remnant undergo 
several bounces, and thereby heat up the central remnant. Note that
for visibility reasons we have restricted  
the color bar to a maximum of 30 MeV in all of the panels. The hot matter from the interface 
region is sheared out and spreads around the central remnant, see column 2, and 
at later times, see column 3, the core is only moderately hot with $T\sim 10$ 
MeV, but engulfed by hot surface layers of $T\sim30$ MeV. While the details 
certainly depend on the astrophysical parameters of the merging binary 
(masses, mass ratio, spins etc.) and the equation of state used, our results
are qualitatively in very good agreement with the results 
found by other thermal EOS implementations \cite{Raithel:2019gws,raithel21a,raithel22,raithel23}
and with simulations using tabulated EOSs, see e.g. \cite{perego19a}.

\subsubsection{Impact of the Dirac effective mass parameters}
To probe uncertainties in the degenerate thermal pressure, we varied 
the parameters controlling the Dirac effective mass $M^*(n)$, namely 
$n_0$ and $\alpha$ in Eq.~\eqref{eq:Meff}. 
Specifically, we considered three representative values of $n_0$
($0.01$, $0.12$, and $0.40$~fm$^{-3}$) and three values of the slope
parameter $\alpha$ ($0.2$, $0.8$, and $1.2$). This set brackets 
the plausible range of thermal responses while keeping the cold 
EOS fixed~\cite{Raithel:2019gws}. The pressures resulting from these choices are shown
as a function of density in Fig.~\ref{fig:thermal-pressure-grid} at $T=1,10,20,30$ MeV and $Y_p=0.1$. 
Each of our nine simulations (three $n_0$ and three $\alpha$ values)
was performed for an equal-mass 
binary with component masses of $1.3\,M_\odot$, starting from 
quasi-circular orbits at an initial coordinate separation of
$45$~km and evolved through merger and $\sim 20$~ms of post-merger
evolution.\\

{\em Density and temperature evolution}\\
The evolution of the minimum lapse function, $\alpha_{\rm min}(t)$, and
the maximum rest-mass density, $\rho_{\rm max}(t)$, can be seen as proxies 
for the overall dynamical evolution. In particular, if the remnant should
undergo collapse towards a BH, $\rho_{\rm max}(t)$ would dramatically 
increase while $\alpha_{\rm min}(t)$ would drop towards zero.
Figure~\ref{fig:alphamin} demonstrates that all models follow the same
qualitative trend which is characteristic for mergers that leave a stable
(or long-lived) remnant: $\alpha_{\rm min}$ decreases sharply at merger,
reaching a minimum, and then it exhibits damped oscillations as the 
remnant settles into what seems a stable configuration. Up to late times 
($\sim 20\,\mathrm{ms}$) the lapse remains stable across all thermal EOS 
parameterizations, none of the models indicates any tendency of collapse.
The corresponding maximum density evolution, shown in
Fig.~\ref{fig:rhomax}, shows strong oscillations during the first few
milliseconds after merger, reflecting the nonlinear core dynamics of the
hypermassive remnant. These oscillations gradually damp, and
$\rho_{\rm max}$ approaches a quasi-stationary value without signs of
runaway growth. This behavior is consistent with the lapse diagnostics
and confirms that the remnant remains supported against collapse for all
thermal models considered here.\\
In  Fig.~\ref{fig:Thot100} we show the peak temperature evolutions 
found as an average over the 100 hottest SPH particles.
During the inspiral, all models remain essentially cold with temperatures below
$T \sim 1\,\mathrm{MeV}$. During the first deep compression in the merger, short-lived 
temperature spikes of $\sim 80$--$100$~MeV appear in the contact layers and 
Kelvin--Helmholtz hotspots, consistent with the illustrative case. The timing 
and amplitude of these peaks vary across parameter choices, with low-$n_0$ models
reaching their maxima later due to a slightly longer inspral time. Within
the next 1--3~ms, the temperatures settle near $T \sim 30$~MeV.\\
Fig.~\ref{fig:EthHigh} shows the thermal energies
in the high-density regions with $\rho>10^{14}\,\mathrm{g\, cm}^{-3}$. During the inspiral 
$E_{\rm th}$ reaches a plateau of a few
$10^{48}\,\mathrm{erg}$. Differences between the models already appear at this
stage, suggesting that variations in the thermal response develop during
the inspiral and may influence the subsequent merger. At merger, the thermal 
energy increases sharply to several $10^{52}\,\mathrm{erg}$ in all simulations. The strongest 
and latest rises occur for the low-$n_0$ models with larger $\alpha$, consistent 
with their delayed merger times. After this phase, the high-density thermal energy
remains largely constant, in line with the presence of a long-lived remnant and
our disregard of neutrino cooling.\\

{\em Impact of the thermal EOS on the inspiral}\\
As shown in Fig.~\ref{fig:Temp_premerger}, the neutron star cores remain essentially cold, 
but the surface temperatures reach values of order $\sim 1$~MeV,
leading to a non-negligible thermal pressure contribution in the outermost
layers. The effect is most pronounced for low values of $n_0$ and larger values
of $\alpha$ (in particular $n_0=0.01$ with $\alpha=0.8$ and $1.2$), for which the
degenerate-matter thermal pressure at $T=1$~MeV is noticeably smaller, as seen
in the top-left panel of Fig.~\ref{fig:thermal-pressure-grid}. This reduced
thermal support effectively results in slightly more compact stars, smaller tidal effects and a
correspondingly delayed merger. The effect is minor for the overall dynamics, 
but it appears in more than one diagnostic and points
to a systematic change in the late-inspiral evolution. To understand it's
origin, we focus on the two parameterizations that yield the earliest and
latest merger, $(n_0,\alpha)=(0.4,0.2)$ and $(0.01,1.2)$, respectively.
Figure~\ref{fig:tidal_elongation} (top) shows a snapshot of the orbital-plane at
$t=6.90\,\mathrm{ms}$. At this time, the $(0.4,0.2)$ case is already strongly
tidally deformed: the high-density region is visibly stretched along the
line connecting the center of masses (COMs) of the two neutron stars, while the
$(0.01,1.2)$ case remains closer to nearly spherical stars. We quantify this trend by measuring, for each star, its extent along three orthogonal directions:
(i) along the line connecting the COMs of the neutron stars, (ii) the direction perpendicular to it within the
orbital plane, and (iii) the direction normal to the orbital plane (the $z$-direction). The elongations
$L_i$ are computed from the region with $\rho\ge 10^{11}\,{\mathrm g\,cm^{-3}}$, and we define the
dimensionless deformation measure 
\be
\eta \equiv \frac{L_{\max}-L_{\min}}{L_{\max}+L_{\min}}.
\ee
The evolution of $\eta(t)$ during inspiral, given in Fig.~\ref{fig:tidal_elongation} (bottom), shows that
the $(0.4,0.2)$ run develops a stronger elongation earlier in time. The associated increase in effective radius enhances the tidal interaction and, given the strong radius dependence of the tidal deformability parameter $\Lambda \sim R^{5}$ (e.g.\ \citealt{hinderer10}), leads to an accelerated inspiral and an earlier merger.\\

{\em Dynamical ejecta}\\
In Fig.~\ref{fig:mej} we show the impact of effective mass parameters  $(n_0,\alpha)$
on the dynamical ejecta. We find that they
have a noticeable effect on the total ejecta mass, with variations of up to
a factor of two across the explored range. To determine whether these
differences also alter the angular or kinematic structure of the ejecta, we
study the mass distribution as a function of polar angle. These profiles
show no significant deviations between parameter choices: all models eject
most mass toward the orbital plane and exhibit a strong decline
toward angles close to the polar axis. This funnel near the poles is consistent
with the expected geometry of neutron-star mergers.
For completeness we also analyze the normalized mass in different velocity bins,
$M_{\rm ej,v}/M_{\rm ej}^{\rm tot}$. The histograms in Fig.~\ref{fig:ejecta_profiles} look nearly identical, indicating that the velocity structure of the dynamical ejecta
is largely insensitive to $(n_0,\alpha)$. Together, these results show that the Dirac-mass parameters mainly rescale the total ejecta mass, while the angular and velocity profiles are governed by the cold EOS part and change little with the thermal response.\\

{\em GWs and post-merger GW spectrum}\\
Figure~\ref{fig:hplus} shows the plus polarization of the $(2,2)$ mode of the
gravitational-wave strain extracted from simulations with
$2\times10^6$ SPH particles for different parameterizations of the effective
mass [cf.\ Eq.~\eqref{eq:Meff}]. During  the early inspiral, the waveforms
overlap nearly perfectly across all thermal EOS extensions, consistent with the
expectation that the bulk inspiral dynamics are governed by the cold EOS.
However, in the late inspiral we observe small but systematic differences in the
phase evolution and in the merger time between different thermal treatments.

After merger, clear differences emerge in the gravitational-wave signal.
The remnant undergoes quasi-periodic oscillations with a rapidly decaying
amplitude, and both the phase evolution and damping time depend on the thermal
EOS parameters $(n_0,\alpha)$, while the overall waveform morphology remains
robust. The corresponding effective amplitude spectra are shown in
Fig.~\ref{fig:psd}, together with the sensitivity curves of Advanced LIGO A+~\cite{aLIGOnoise}, Einstein telescope (ET)~\cite{Maggiore:2019uih}, and
Cosmic Explorer (CE)~\cite{Reitze:2019iox}. All models exhibit a dominant spectral peak in the
$2$--$4$~kHz range, associated with the fundamental quadrupolar oscillation mode
of the hypermassive remnant. The location of this peak shifts systematically with the effective mass
parameters, with typical variations of order $\Delta f \sim 150$~Hz across the
explored thermal EOS parameter space at fixed cold EOS. Subdominant spectral
features, related to secondary mode couplings, also show parameter dependence,
although at lower amplitudes. 

Whether such thermal effects can be disentangled observationally remains to be
seen, but it will definitely be challenging since the effects can compete with other
physical processes related to magnetic field evolution and neutrino emission. 
Future gravitational-wave detectors, including Advanced LIGO A+ and
third-generation facilities, are expected to access the dominant post-merger
peak frequency. In combination with inspiral constraints on the cold EOS,
post-merger spectroscopy may therefore provide a path toward probing the
finite-temperature sector of dense matter. A quantitative assessment will
require higher-resolution simulations, improved waveform modeling, and
dedicated statistical analyses.

\section{Conclusion}
\label{sec:summary}

In this paper we have presented a substantial methodological update of the
Lagrangian numerical relativity code \texttt{SPHINCS\_BSSN}.
We abandoned our earlier convention of measuring all energies in units of
the baryon rest-mass energy and now use  standard numerical relativity
units. More importantly, we have implemented constraint-damping terms 
in the BSSN spacetime evolution, closely following recent developments proposed in an Eulerian context \cite{etienne24}. We demonstrated that this modification reduces Hamiltonian-constraint 
violations by more than an order of magnitude at essentially no additional computational cost.
Most importantly from a physical perspective, we implemented a
finite-temperature extension of cold equations of state based on Fermi liquid
theory, following the framework introduced by Raithel et al. \cite{Raithel:2019gws,raithel20}.
This approach replaces the commonly used ideal gas prescription with a
physics-based treatment of thermal energy and pressure as a function of density, temperature, and arbitrary proton fraction,
while retaining full analytic control and avoiding the computational overhead
and potential robustness issues associated with tabulated EOSs.

Using a set of fully general relativistic merger simulations, we demonstrated
that the new thermal EOS implementation produces temperature distributions,
and remnant structures in good qualitative agreement with
previous studies employing tabulated finite-temperature EOSs. 
During the inspiral phase, we find that the neutron-star cores remain
effectively cold, while tidal interactions heat the low-density surface layers
to temperatures of order $\sim 1$~MeV. In the present simulations, this surface
heating leads to small differences in the late-inspiral gravitational-wave phase
evolution and merger time. During the merger,
peak temperatures locally reach close to 100~MeV  during the strongest compression phase, but
the remnant subsequently settles into a new configuration with a 
moderately hot core ($\sim 10$ MeV) surrounded by hotter ($\sim 30$ MeV) 
surface layers, consistent with results reported in the literature.

By systematically varying the parameters that control the density dependence
of the Dirac effective mass, we explored uncertainties in the degenerate thermal
sector. We found that these parameters mainly rescale the total thermal energy
and the amount of dynamical ejecta, which can vary by up to a factor of two across
the explored range. In contrast, the angular and velocity structure of the ejecta
remain essentially unchanged.
Thermal effects leave a visible imprint on the post-merger phase.
In particular, we find systematic shifts of order $\sim 150$~Hz in the dominant
post-merger spectral peak when varying the thermal EOS parameters at fixed cold
EOS. While these shifts are subdominant compared to changes induced by different
cold EOSs, it remains to be seen whether they can be measured observationally.
Future gravitational-wave detectors, such as Advanced LIGO A+ and third-generation
facilities, are expected to access the post-merger peak frequency, providing a
potential avenue for confronting such thermal effects with observations.\\

\section*{Acknowledgements}
BB and SR acknowledge the support from the Knut and Alice Wallenberg Foundation 
under grant Dnr.~KAW~2019.0112, the Deutsche 
Forschungsgemeinschaft (DFG, German Research Foundation) under 
Germany's Excellence Strategy – EXC~2121 ``Quantum Universe'' –
390833306. BB was further supported by the Alexander von Humboldt Foundation through a Humboldt Research Fellowship for Postdoctoral Researchers. SR has additionally been supported by the Swedish Research Council (VR) under grant number 2020-05044, by the research environment grant “Gravitational Radiation and Electromagnetic Astrophysical Transients” (GREAT) funded by the Swedish Research Council (VR) under Dnr 2016-06012,  and by the European Research Council (ERC) Advanced Grant INSPIRATION under the European Union’s Horizon 2020 research and innovation programme (Grant agreement No. 101053985) which also supports LS. \\
The calculations were performed in part at the NHR Center NHR@ZIB, jointly supported
by the Federal Ministry of Education and Research and the state
governments participating in the NHR (www.nhr-verein.de/unsere-
partner), at the SUNRISE HPC facility supported by the Technical
Division at the Department of Physics, Stockholm University, and
on the HUMMEL2 cluster funded by the Deutsche Forschungsgemeinschaft (498394658). Special thanks go to Mikica Kocic (SU),
Thomas Orgis and Hinnerk Stüben (both UHH) for their excellent
support. Portions of this research were conducted with high performance computational resources provided by the Louisiana Optical Network Infrastructure (http://www.loni.org).

\bibliography{mybiblio,astro_SKR.bib}

\end{document}